# Windtalking Computers:
# Frequency Normalization, Binary Coding Systems, and Encryption

Givon Zirkind
B.Sc., Computer Science, Touro College; M.Sc. Computer Science,
Fairleigh Dickinson University
givonz@hotmail.com

## ABSTRACT

The goal of this paper is to discuss the application of known techniques, knowledge and technology in a novel way, to encrypt computer and non-computer data. There are two distinct and separate methods presented in this paper.

Method 1: Alter the symbol set of the language by adding additional redundant symbols for frequent symbols. This will reduce the high frequency of more commonly used symbols. Hence, frequency analysis upon ciphertext will not be possible. Hence, neither will decryption be possible.

Method 2: [A prerequisite to understanding this method is to understand that there is a difference between a binary representation and base 2.] To-date most computers use the binary base 2 ($base_2$) and most encryption systems use ciphering and/or an encryption algorithm, to convert data into a secret message. The method of having the computer "speak another secret language" as used in human military secret communications has never been imitated. The author presents the theory and several possible implementations of a method for computers for secret communications analogous to human beings using a secret language or; speaking multiple languages. This is done by using a binary base other than base 2. Ex. Fibonacci, Phi or Prime.

In addition, steganography may be used for creating alternate binary bases. This has no mathematical resolution if implemented with randomness.

This kind of encryption scheme proposed significantly increases the complexity of and the effort needed for, decryption. First the binary base must be known. Only then, can decryption begin.

This kind of encryption also breaks the transitivity of plaintext-codebook-binary. Or, the correlation between letters-ASCII-base2. With this transitivity broken, decryption is logically not possible. (This is discussed and explained in detail.)

Coupled together with encrypting the plaintext, binary encryption makes decryption uncrackable, produces false positives—information theoretic secure, and requires much more computing power to resolve than is currently used in brute force decryptions. Hence, my assertion that these combination of methods are computationally secure—impervious to brute force.

As every methodology has its drawbacks, so too, the proposed system has its drawbacks. It is not as compressed as a $base_2$ would be. (Similar to adding random padding to the encryption.) However, this is manageable and acceptable, if the goal is very strong encryption:

At least two of the general methods and their various implementations herein proposed are not decryptable by method – uncrackable – by conventional, statistical means.

Specifically:

1. Creation of new symbol sets is used to alter and confound the natural symbol frequency.
2. Also, alternate binary encryptions other than binary base 2 are used.

Using alternate binary encryptions lend easily to the creation of new symbol sets and the confounding frequency analysis.

## Categories and Subject Descriptors

D.2.11 [**Software**]: Software Architectures – *Data abstraction*
E.0 [**Data**]: General
E.3 [**Data**]: Encryption
E.3 [**Data**]: Encryption – *Code Breaking*
E.m [**Data**]: Miscellaneous
F.2.0 [**Theory of Computation**]: Analysis Of Algorithms And Problem Complexity – *General*
F.2.1 [**Theory of Computation**]: Analysis Of Algorithms And Problem Complexity – Numerical Algorithms and Problems – *Number-theoretic computations*
F.2.2 [**Theory of Computation**]: Analysis Of Algorithms And Problem Complexity – Nonnumerical Algorithms and Problems – *Pattern matching*
F.2.m [**Theory of Computation**]: Miscellaneous
H.0 [**Information Systems**]: General
H.1.0 [**Information Systems**]: Models and Principles – General

H.1.1 [**Information Systems**]: Models and Principles – Systems and Information Theory – *Information Theory*

# 1. INTRODUCTION

## 1.1 Revisions:

This article was originally published in April 2008. The article did not pass peer review. I self published on the web. Also, after years of not being able to pass peer review in academic circles; I sought support from the hacker community (white & gray hat only) to attempt to crack a sample encrypt. Possibly, by setting up a distributed net to attempt a brute force decrypt. I did not say how to attempt or program that brute force decrypt. (My methods are known and available from the ACM archive, part of the Cornell archive.) Over time, I have gotten several comments about the article. These comments can be divided into categories. Also, there are some comments that reoccur. This revised article is to address the category of statements and the commonly reoccurring statements. A lot of history and anecdotes have been removed from this article. Many examples have been deleted from this article. The Abstract has been revised, reformatted and has many additions. A lot of information has been removed in the hopes of simplification and clarity. The original can be found at http://arxiv.org/abs/0912.4080

I fully understand the resistance to someone making the claim to having made an advance in the field of encryption. I fully understand the oft asserted claim of making an encryption that is uncrackable. Which is why I always preface my emails and articles, with a brief bio, of my academic background as well as the fact that I own a patent in data compression. Data compression inherently involves encryption. Along with the achievement of having solved a compression with a 20 year standing engineering prize. Also, since the claim is tall, I had all my colleagues review the article before submitting to any journal.

Even thought the claim is tall, intellectual probity demands actually reading an article before assuming it is bunk. Real refutations require stating the premise and steps of the logical proof, THEN; demonstrating that either the premise or one of the steps of the logical proof are false. Detractors that don't do that, are not valid detractors. Detractors who say, as some PhDs in cryptography have said, 'Your math is potentially ok.' But, refuse to accept my conclusions, do not have probity. [Or, perhaps they are government hacks trying to dissimulate? So, more uncrackable encryption methods will not be available to the public?]

I readily admit cryptography is not my forte. At the same time, I am not ignorant of the subject either.

If a scientist is to have true probity, he has to actually read the article before rejecting the method. I have received many offhanded, spurious dismissals. No one can read a technical article of this length in 5 minutes. I accept those offhanded dismissals as a fact of life. Arrogance. Intellectual arrogance. Professorial arrogance. Scorn of a presumptuous neophyte. Not as a real refutation.

The honest and good say they are not interested in evaluating such a tall claim and are not going to read such a paper. – I can accept and appreciate that.

I will note a significant aspect of the rejections so far. The more academic or government related the reviewer the less academic and more personal of an attack, the rejection is. Many prejudicial assumptions are made—of me personally or my claims. The ultimate of course is the exasperated, "it doesn't make sense" without explanation. Remember, especially in the U.S., schools and research centers are heavily government funded, especially in the field of cybersecurity. The less government affiliated the reviewer – business person (sysadmin or security professional) or tech writer or hacker (or the term cracker if you prefer as a non-malicious curious individual) – the more they understand.

Also, by simple comments and questions, I can tell:

1. If someone actually read the article. Any reviewer that discusses and rejects my method—singular—has either not read the article, nor understood it. All except one reviewer [Travis H.] has mentioned just one method. *That* reviewer did NOT give an offhanded dismissal to my claims. Not to say he agrees. But to say, he warrants reading the article and contemplating the processes discussed.

2. If someone understands the heart of the linguistics and math involved or not.

First, I describe two basic categories of encryption. That is mentioned in the title. Frequency normalization is the first. I devote an entire section to this one process. Even laymen get that.

<u>First Method:</u> Frequency normalization is a common mathematical term also employed in physic, probability and other fields. As the basis of decryption is to look for the natural frequency of letters in a ciphertext—basing an encryption process upon the efficacy of confounding the natural frequency of letters is something I will not belabor an explanation. This is fundamental to cryptography.

Once the frequency has been normalized, then standard encryption techniques are applied. The binary techniques here may be applied also for good measure. But once the frequency has been normalized; and then the plaintext encrypted; the frequency analysis commonly employed in decryption will be of no avail.

Only one person [from the Cryptography listed in the acknowledgements] commented, so far has asked me why you must change the symbol set. At least he asked the right question. The answer is the heart of the *first* encryption process. Without changing the symbol set, confounding the natural frequency is not possible. (Perhaps reading my article, [ZIR01] on data compression might help explain why. Read the section that describes, why serial compression will not gain in compression.)

This method of enciphering or encoding or encryption (each word has its own technical meaning is this list) by altering the symbol set has never been used in the field of encryption. All the reviewers –like one reviewer for the American Mathematical Society (AMS)—who said, 'No one has ever done it this way before.' Only confirms my assertion of the novelty of the encryption methods I propose.

If no other encryption would be employed together with frequency normalization, then decryption would involve: guessing which sets of letters represent the original letter; then observing if the words created are in the dictionary and if the entire message makes sense.

This decryption should be, I believe, information theoretic. There would be false positives. While I have not calculated the probabilities or statistics, the simple number of combinations would seem to me, to generate some false positives.

Decrypting this method alone may or may not be computationally secure. This method alone may not be impervious to brute force. [A human could easily decrypt this. A computer would require lots of guessing and may not be able to guess right.] But, coupled with any other encryption—of any kind—even the simplest—it would be logically "uncrackable" and beyond brute force decryption.

<u>Second method:</u> Involves breaking the transitivity of letter—codebook—actual binary representation. Commonly, letter—ASCII-base 2. [Explained in depth in the article.] This is done with a combination of two techniques:

1. A choice of encryption of the letter or using a different codebook (reference chart, ex. EBCDIC instead of ASCII) or both changing the codebook along with encrypting the letter.
2. Encrypting the binary.

Encrypting the binary must be done in a way that has no mathematical relationship to the encryption of the codebook. Otherwise, transitivity is maintained. This is explained in depth in the article.

Encrypting the binary is done either with steganography or; by using other natural bases –other than base 2—which can be represented in binary. (Ex. Fibonacci and Phinary) This is discussed in depth in the article.

[Steganography in computers is not new. [JOH001] People have been embedding messages in the bytes of JPEGs and pictures for decades. The particular technique of computer steganography I propose in my examples may or may not have been used before. I have not heard of its use before. I have not researched its existence or use. I will leave the priority usage to computer historians to judge.]

One very intellectual commentator [Travis H.] stated that without randomness, the steganography is crackable.

True. While randomness can be implemented into the technique, I intentionally leave out how to implement randomness in this technique. Suffice to say, randomness can be built into the steganography. However, in my original article I discuss the subject when discussing using big and little endian as steganographic techniques. Also, I did mention "changing the masks". Bear this in mind when reading that section.

Some individuals who reviewed the article, told me, that unless randomness was introduced—which must be done by padding—the method would be crackable. The original article discusses this too. (That section has been preserved as Section 7, "Infinite Combinations", in this article.) That when encrypting the binary from base 2 to another binary base, padding is a natural consequence. Padding is also a natural consequence of the steganography.

It is obvious if you look at the pictures in the article, that the steganography pads the ciphertext.

## 1.2    Amended From Original Introduction:

The order of several sections have been change so that the presentation of methods will match the order of the methods listed in the title.

Reordering the material included renumbering the tables. Although the table references were proof read, there may be some errata in the references.

Two major section headings have been added to indicate where those methods are being discussed.

A section has been added explaining with changing the symbol set is not decryptable by brute force.

In a previous published article, I mentioned parenthetically, in a discussion of compression and encryption, that the exchange of symbol sets (ciphering) does not alter the frequency of the symbols [ZIR01]. Hence, the original symbol set can always be ascertained from an enciphered text [KAH05]. Intrinsically altering the symbol set is one way to truly encrypt a linguistic message. This article discusses the process of altering the linguistic symbol set and the impact that altering the linguistic symbol set has upon encryption. In addition, this article discusses how to encrypt the binary numeric representation, exclusive of the alphabetic encryption and; the impact binary numeric encryption has upon decryption. I.e. Producing a very strong encryption that can not be decrypted, certainly not methodically by process.

[The linguistic symbol set would be letters as opposed to a numeric symbol set of numerals.]

I have made it (non-decryptable or uncrackable encryption) much easier and more practical than one time key encryption.

What I have done, improved upon, is, that practically speaking, one time key encryption can now be decoded by guessing. That is what brute force methods do. I have increased the guessing possibilities astronomically. Whether supercomputers or distributed networks can crack these encryption methods, remains to be seen.

On another level, what I have done is to advance the level of intelligence in language. There can be no intelligence without language. [WHO01] Language inherently involves intelligence and the expression of ideas. [WHO01] One of the chief methods that language employs in expressing ideas and using intelligence is comparison and categorization. [WHO01]. Language involves the expression of covert as well as overt ideas and concepts. [WHO01] Which, is done by categorization. [WHO01] Encryption, in a regard, involves a high level of intelligence, linguistically speaking, by making extremely covert categories. The reverse, decryption, involves a high level of intelligence, by recognizing the hidden patterns of language in a seemingly mass of chaos. What I have done, is to raise our level of intelligence to the point, that even though there is an overt chaotic mass, in which we can not see the intelligence, at least (upon generation) we can categorize the different, non-interpretable (non-decryptable) chaotic masses containing intelligence.

[One professor commentator said that he has no idea what I am talking about. But, suggested that perhaps I was talking about using chaos theory for encryption. His comment implied been there done that. To be clear, although I did use the word chaos in the previous paragraph, the methods I propose, do not, intentionally have anything to do with chaos theory. This all about logic, math and linguistics.]

This is the indicator to understanding this paper. When one reads this presentation, one should understand, that there are many examples of several new methods. If one sees just one method, one has to study the paper further. [I have adjusted the titles to make this clear.]

Writing is a relationship and correspondence between sounds, lexicons and written symbols. The permutations and combinatorics of all the possible sounds, to make as many words as we need, is a necessity of communication. Those permutations and combinations must be defined and limited by the natural phenomenon of what the mouth can utter and the ear can hear. Permutations and combinatorics are math. A lexicon and grammar relate objects (words). Relationships are logic. So, math and logic are an intrinsic part of communication. When you can no longer do the math and; there are no relationships (It's illogical.); then, there can be no communication. That is encryption.

## 1.3 Responses to Additional Comments

→ At this point it is fitting to answer a common comment about my work, "Why would I want to use non-decryptable encryption? Why would I want to encrypt something that I could not decrypt?" To take this comment seriously, it is a gross misunderstanding of encryption and; an ignorance of the fundamental concept of encryption. Any encryption or enciphering or encoding (these are three distinct technical terms chosen specifically for a reason), is designed to be non-decryptable—to the uninitiated. But the sender and receiver will have the keys and methods to decrypt.

Obviously, I did not intend to produce total gibberish that can never be understood again.

Obviously, I did not intend that there should be no reverse process for decryption.

What I intended by "non-decryptable by method" is "uncrackable".

What I intended is, that if you do not have the keys and you do not know the exact methods of encryption (the alpha & binary), then, you can not decrypt the message. Even with the advanced computing power of the today. Your information is secure. That is what encryption is all about.

→ One comment and misconception that constantly arises is a misunderstanding of terms. In the original paper I used the term "non-decryptable by method". Apparently, this is an old term, no longer used. The term "uncrackable" or "unbreakable" seems more appropriate.

[Jeremy Stanley, commented on Jack Lloyd's Cryptography list, that he was not able to find the expression "non-decryptable by method" in literature. I responded, that I had used the term from the original patent for one time key pad from the 1930s. So, I have gone with the suggestion "uncrackable".]

He also asserts that this would seem an exaggeration. Because of brute force. True. But, it is no more an exaggeration than the ability to decrypt one time key pad. And, if you need more computing power than currently available—a common fallback for many encryption schemes—than this new method is no exaggeration. That is why we constantly hunt for bigger & bigger prime numbers. To require more computing power to decrypt, than is currently available.

→ As for not having passed peer review in the past:

The so-called refutations lack any logical and mathematical refutation.

One absurd declaimer said, that my math is potentially ok but he doesn't know what I am talking about. A clear contradiction. If my math is ok, then, I am right? Ne c'est pas?

All these so-called refutations say that encryption was never done this way before. So my claim of novel techniques is certainly valid.

→ Previously, I had a lengthy discussion in the introduction about releasing the genie from the bottle. That in our age of terrorism, cybercrime, cyberwarfare, would it be prudent –if I really had found a new uncrackable, unbreakable encryption— would it be prudent to publish a paper about it? Well, if the academic community and tremendously funded government agencies entrusted to this science say I am full of hokum, then the release of this information and techniques is of no moment—certainly of no moment to governments.

→ Here I will add some history, that I did not mention in the original version. This history is now appropriate:

When David Kahn, a newpaper reporter, wrote his "History of Cryptography", an excellent work for an amateur and a good primer for encryption, the NSA (National Security Agency, U.S.) maligned him as an amateur and his work as incorrect. If so, why was the NSA pressuring the publisher not to publish his book? When the NSA asked for the removal of certain information which they considered sensitive—which was public record anyway—the author simply agreed. [BAM01] The NSA could have chosen the correct, civil and polite method. It didn't need to be nasty, maliciously degrading and maliciously discrediting.

I believe we are witnessing a repeat performance by the NSA and other government types of attacking law abiding citizens, who are not part of the government cryptographic apparatus. Today, this runs counter to the federal law in the U.S., the Cybersecurity Act. It would behoove the U.S. government and other governments to work with intelligent and talented, law abiding private individuals. Rather than, attacking and harassing them.

For this reason, I have eliminated any apologies for unleashing a new "uncrackable" encryption method. I intentionally submitted prior versions of my article to journals inside the U.S. only. (Where was living at the time.) However: If the academic community can not understand my presentation; then I am not to blame for disseminating gibberish. If the government with all their experts, can not understand my presentation, their loss. I was intentionally obfuscating, but included enough for intelligent, open minded people knowledgeable in the field to get it. If hackers got it, while the government and academics didn't; the governmentals and academics need to rethink their position and academic status. If the hackers have already understood, then the genie is already out of the bottle. This article, written with clarity intended, should not cause any clamor about unleashing a new "unbreakable" encryption method.

I am reminded of the anecdote of the man who's car has engine trouble. He brings his car to his mechanic. His mechanic can't find the problem. His car breaks down on the highway. He goes to get help. By the time a tow truck comes to tow his car to a garage, his car is stolen. What the mechanics couldn't fix for weeks, the thieves could fix in a minutes!

Had the government chosen the correct path of classification and restriction of technology, as a law abiding citizen I could have accepted that. Although, with all the cyrberattacks and cybercrime, in my opinion, John Q. Public deserves better encryption and security than the government is allowing or; industry is providing.

Since, there are issues to implementation, which I have not explained how to overcome, that is sufficient for my conscience for not having released the genie from the bottle. Because, no one who understands the process, so far, has been able to overcome the obstacles to implementation. (Also, omitting several equations that I had to conceive as well as an integral mathematical concept.)

## 2. FREQUENCY NORMALIZATION
## 2.1 CHANGING THE SYMBOL SET

As I stated and partially explained above as well as in a previous article [ZIR01], what is necessary for compression and encryption is to change the symbol set. For example, if the English text were replaced with the international phonetic alphabet, there would be more letters altogether and less of some regularly written letters. I.e. The dipthong "ch" would be replaced by one symbol. Now this new symbol appears, with its own frequency. And, the frequencies of 'c' and 'h' have been changed. Because, 'c' and 'h' no longer appear where 'ch' appears – the dipthong has been replaced by a new symbol.[1]

This substitution technique is also well known in cryptography. It does complicate matters. [KAH12] [KAH13] [VAU03]

Applying this knowledge to the premise of this paper: The properties of the binary symbol set are altered so that the alphabetic symbol set is altered in a way that "normalizes" the frequency of the letters. This is a completely novel approach to encryption. This encryption method does not use a key per se; although one must have the translation table.

What I am proposing is: Add extraneous, "useless" binary numbers to create extra symbols to alter the frequency. Then encryption methods are used on the letters of the original text, to produce a ciphered text that includes new identity letters which do not have the frequency analysis of English. This would be sufficient to not be decryptable. But, in addition to altering the frequency, the ciphered text is (for example) then translated to binary base 2 which is enciphered in base Fibonacci. With extraneous numbers?! – Then, the encryption is very strong and not decipherable!

Example:

Returning to the "The quick brown fox jumped lazily over the sleepy dog." Let's say, I replace some of the 'e's with another, new, symbol. E.g. "The quick brown fox jumpφd lazily over thφ sleφpy dog." I have achieved several things. Most importantly, the frequency analysis is now confounded. The letter 'e' is no longer the most frequent letter. If I would now engage in some kind of enciphering, there is no way to get back to the original message by frequency analysis. [Nor, is there any other method that I can think of, that will reverse the process

---

[1] See citations [WIK10] & [WIK11] for more information on the International Phonetic Alphabet (IPA), as well as a link for a chart of the IPA.

without knowing the codebook encryption process. (The extra symbols and what they replace.)]

In a sense, this kind of encoding is a CAPTCHA. When a human sees this encryption, a human knows that it is seeing an encryption and; what is encrypted. But, a computer, using dictionary attacks, by trying to match the letter patterns of words in a dictionary will not. The computer can only offer the human a possible, but not definite, decryption.

Notice how "the" and "thφ" are now 2 different words. The confusion begins.

If I follow the procedure through, and apply the technique to the 2nd most frequent letter, "O", then the sentence now looks like: "The quick brown fψx jumpφd lazily over thφ sleφpy dψg."

Look at Tables 1A, 1B & 1C, below comparing the 3 frequencies for the 3 different enciphering methods:

| Frequency for unaltered text. | | |
|---|---|---|
| Letter | Tally | Frequency |
| E | 6 | 6/44 = 13.6% |
| O | 4 | 4/44 = 9.0% |
| L | 3 | 3/44 = 6.8% |
| D | 2 | 2/44 = 2.5% |
| H | 2 | 2/44 = 2.5% |
| I | 2 | 2/44 = 2.5% |
| R | 2 | 2/44 = 2.5% |
| T | 2 | 2/44 = 2.5% |
| U | 2 | 2/44 = 2.5% |
| A | 1 | 1/44 = 2.3% |
| B | 1 | 1/44 = 2.3% |
| C | 1 | 1/44 = 2.3% |
| F | 1 | 1/44 = 2.3% |
| G | 1 | 1/44 = 2.3% |
| J | 1 | 1/44 = 2.3% |
| K | 1 | 1/44 = 2.3% |
| M | 1 | 1/44 = 2.3% |
| N | 1 | 1/44 = 2.3% |
| P | 1 | 1/44 = 2.3% |
| Q | 1 | 1/44 = 2.3% |
| S | 1 | 1/44 = 2.3% |
| V | 1 | 1/44 = 2.3% |
| W | 1 | 1/44 = 2.3% |
| X | 1 | 1/44 = 2.3% |
| Y | 1 | 1/44 = 2.3% |
| Z | 1 | 1/44 = 2.3% |

Table 1A

| Frequency with one additional new letter. | | |
| --- | --- | --- |
| Letter | Tally | Frequency |
| O | 4 | 4/44 = 9.0% |
| E | 3 | 3/44 = 6.8% |
| L | 3 | 3/44 = 6.8% |
| D | 2 | 2/44 = 2.5% |
| H | 2 | 2/44 = 2.5% |
| I | 2 | 2/44 = 2.5% |
| R | 2 | 2/44 = 2.5% |
| T | 2 | 2/44 = 2.5% |
| U | 2 | 2/44 = 2.5% |
| A | 1 | 1/44 = 2.3% |
| B | 1 | 1/44 = 2.3% |
| C | 1 | 1/44 = 2.3% |
| F | 1 | 1/44 = 2.3% |
| G | 1 | 1/44 = 2.3% |
| J | 1 | 1/44 = 2.3% |
| K | 1 | 1/44 = 2.3% |
| M | 1 | 1/44 = 2.3% |
| N | 1 | 1/44 = 2.3% |
| P | 1 | 1/44 = 2.3% |
| Q | 1 | 1/44 = 2.3% |
| S | 1 | 1/44 = 2.3% |
| V | 1 | 1/44 = 2.3% |
| W | 1 | 1/44 = 2.3% |
| X | 1 | 1/44 = 2.3% |
| Y | 1 | 1/44 = 2.3% |
| Z | 1 | 1/44 = 2.3% |

Table 1B

| Frequency with two additional new letters. | | |
| --- | --- | --- |
| Letter | Tally | Frequency |
| E | 3 | 6/44 = 6.8% |
| L | 3 | 3/44 = 6.8% |
| D | 2 | 2/44 = 2.5% |
| H | 2 | 2/44 = 2.5% |
| I | 2 | 2/44 = 2.5% |
| O | 2 | 2/44 = 2.5% |
| R | 2 | 2/44 = 2.5% |
| T | 2 | 2/44 = 2.5% |
| U | 2 | 2/44 = 2.5% |
| A | 1 | 1/44 = 2.3% |
| B | 1 | 1/44 = 2.3% |
| C | 1 | 1/44 = 2.3% |
| F | 1 | 1/44 = 2.3% |
| G | 1 | 1/44 = 2.3% |
| J | 1 | 1/44 = 2.3% |
| K | 1 | 1/44 = 2.3% |
| M | 1 | 1/44 = 2.3% |
| N | 1 | 1/44 = 2.3% |
| P | 1 | 1/44 = 2.3% |
| Q | 1 | 1/44 = 2.3% |
| S | 1 | 1/44 = 2.3% |
| V | 1 | 1/44 = 2.3% |
| W | 1 | 1/44 = 2.3% |
| X | 1 | 1/44 = 2.3% |
| Y | 1 | 1/44 = 2.3% |
| Z | 1 | 1/44 = 2.3% |

Table 1C

Analyzing the data in Table 1, we see that with only one new symbol, 'e' is no longer the most common letter. If we use two new symbols, 'e' is tied in first place with another letter and; 'o', a very frequent number, becomes an ordinary number.

The additional letters are not shown. The additional letters have the same frequency as the letters they replace. Although, we could do this differently so that one new twin letter has a higher frequency, than its additional twin letter. This means, for the most frequently occurring letter, there is at least two, if not three letters tied for 1st place.

Would a decryption method start by guessing that some combination of the most frequent letters is 'e'? Then, what?

And, if I follow through with the identity replacement procedure, and apply the technique to punctuation marks, like spacing, then the sentence now looks like: "Theωquick brownωfψx jumpφdωlazily overωthφ sleφpyωdψg." And, if you do not know where words begin and end, you are missing a big clue in deciphering. Human beings intuitively know, when deciphering, if they are seeing words or not and where spaces, breaks between words, should go. Computers have to do a dictionary comparison on text and subtext: Guess.

A dictionary comparison creates longer and longer cumulative length strings, compares them with EVERY word in the dictionary and; then, decides if a possible word or word, has been located and a new string should be started for analysis.

E.g. Using the example above, a dictionary comparison would be done like this:

The secret message is:
"Thequickbrownfoxjumpedlazilyoverthesleepydog."

"T" – is "T" a word? No. Add a letter.
"Th" – is "Th" a word? No. Add a letter.
"The" – is "The" a word? Yes. Record first word. Start new word.
"q" – is "q" a word? No. Add a letter.
"qu" – is "qu" a word? No. Add a letter.
"quic" – is "qui" a word? No. Add a letter.
"quic" – is "quic" a word? No. Add a letter.
"quick" – is "quick" a word? Yes. Record second word. Start new word.

Here we can see the difficulties in this analysis and the propensity for errors. After having discovered the first four words, "the", "quick", "brown", "fox"; the next word is "jump". Then the analysis becomes unclear. Should there be a look ahead? If so, how many letters to look ahead? Is it "over" or "overt"? It could it be "overt he" instead of "over the". Is slang included or not? I.e. Is "bro" a word?

True, there are sophisticated dictionary algorithms. But, the point is that by simply removing the space between words creates a big obstacle for a computer to decipher. -- Consider this, removing a space is alteration of the symbol set. There is now, one less symbol. If half the 'e's in the sentence had been removed, it would be a big alteration.

While computers can figure out, if the string of words deciphered actually makes sense as a sentence in English; this is not accurate and is not as sophisticated as a human.

## 2.2 Frequency Normalization

In addition, to just replacing every other letter, knowing the letters' frequencies from tables, etc.; with little effort, one could parse any given input text, tally all the letters, calculate the frequency of each letter, and then; create a sufficient amount of new symbols – identity symbols – to "normalize" enough of the letters to make frequency analysis impossible! All that would be required is to randomly replace the given letters with their corresponding identity letters.

In fact, a very similar method is done in JPEG, with Huffman coding. [MIA01] [PEN01] The JPEG standard has a feature, to use Huffman coding. Huffman coding is the most compact binary encoding. [HUF01] What Huffman coding does, is tally all the symbols, and assigns the shortest code symbol to the most frequent symbol to be coded. I.e. The most frequent letter, for example, will be replaced with a '1'. But, Huffman coding requires parsing the data prior to encoding. This is necessary to develop the code table. Huffman encoding produces a table which must be transmitted for decoding. Also, in JPEG, each new section of data – image frame, will have its own Huffman table as the frequency of the symbols will change in each frame. For example, in one frame, you may have hundreds of red pixels – which will be assigned a '1'. And, only a few white pixels, which may be assigned a '1111'. But, in the next frame, there may be hundreds of white pixels – which will then be assigned a '1'. And, the few red pixels, which will be assigned a '111'.

Since JPEG is not a secret encryption method, there is no concern about the code table being captured and the data being decrypted. But, with secret messaging, it will be very important to protect the code tables. – A disadvantage. But, a common concern in secret encryption and a concern that is no more significant than protecting an encryption key.

[In [ZIR01] I discuss JPEG and Huffman encryption in depth.]

Following are possible implementations of this theory:

## 2.3 Base 2 Binary Frequency Normalization

Even with regular binary, not all the symbols of the ASCII table are used. There are enough symbols left over for graphics and alternate (other language) alphabets (e.g. Greek). What if, for example, the symbol 253 was also used to represent the letter 'e', along with symbol 69?

Following such an insertion of additional identity symbols, with one simple binary encryption, the entire frequency analysis is irrevocably confounded! No complex key is needed to make and keep the message secret!

## 2.4 Fibonacci Frequency Normalization

Using a more simple and obvious example first: Any positive integer less than any given Fibonacci number can be had, by adding up some set of the previous Fibonacci numbers. For example, the number 6 is not in the Fibonacci sequence, but, it could be composed of 1+5; or 1+2+3. In fact, adhering strictly to the Fibonacci sequence; {0, 1, 1, 2, 3, 5….} – the number (big endian) "010001$_{FIB}$" (1+5) is not the same as "001001$_{FIB}$" (1+5)! It would be a trivial matter to create extra symbols that referred to the same letter.

There is no one to one correspondence between positive integers and Fibonacci numbers. – This is one of the beauties of base Fibonacci. This property alone, becomes very useful for implementing a frequency normalization encryption scheme as herein discussed, without any serious extra effort encoding! Simply pick different sets of numbers to express the ASCII numbers for any letter – and alternate the set you use! Most letters will have corresponding ASCII numbers with multiple corresponding Fibonacci sets.

In addition, if there is no frequency analysis to be had, there is also no way of knowing which binary system is being used! For, even if we assume, that the binary system was encrypted, and; that it would be possible to determine which binary system was being used; by assuming that we could backtrack from a valid language letter frequency analysis; after frequency normalization of the language letter symbol set, that clue – the language letter symbol frequency – no longer exists. After frequency normalization, any binary system that does produce a proper frequency analysis is producing a false positive! You would not know, that the frequency analysis, did indeed provide a link to the correct binary system or not.

## 2.5 An Example of Frequency Normalization

To give a real example, with a larger sample than one sentence: Let's use the abstract of this paper which is only one paragraph. When that abstract was written, there was no intention of choosing words so that a special frequency would occur or; that every letter of the alphabet would be used or; that any letter would be excluded. It was just written to express a thought. The text was not adjusted in any way to facilitate encryption or be an example of any kind. I am positive very similar results will occur if I choose to work on the King James' Version of the Bible or any other text for that matter.

Look at Table 2. It is divided into three sub-tables A, B & C. Each sub-table has three columns. The leftmost column contains the letters of the alphabet. The middle column contains that letter's tally. The rightmost column expresses that letter's tally as a percentage of the total number of letters in the paragraph.

- The leftmost sub-table is sorted alphabetically.
- The middle table is sorted by percentage, in descending order. I.e. Most frequent letter first. Followed by the next most frequent letter. Etc.
- The rightmost sub-table shows what the percentages would be, if identities had been issued and frequency normalization applied.
    - The additional new identity symbols are not displayed.
    - The frequencies of the new symbols are assumed to be the same as the letters that they replace.
    - The encryption method applied:
        - Adding three additional symbols to the most frequent letter. Its frequency is altered as it appears in the third sub-table.
        - Adding two additional symbols to each of the next, top 9 most frequent letters. Their frequencies are also altered as they appear in the third sub-table.
    - Adding new symbols to a given letter, in effect divides the tally of that given letter.
    - The letter 'e', having such a high frequency, was given three new symbols. So, 'e', goes from a frequency of 12% to 3% (12%/4=3%).

| Alphabetized Frequency Analysis | | | Frequency Analysis Sorted by Numeric Value | | | Frequency Analysis With Identity Letters Added; Sorted by Numeric Value | | |
|---|---|---|---|---|---|---|---|---|
| A | 67 | 8.5% | E | 96 | 12.1% | E | 96 | 3.0% |
| B | 10 | 1.3% | T | 75 | 9.5% | T | 75 | 3.2% |
| C | 36 | 4.5% | O | 71 | 9.0% | O | 71 | 3.0% |
| D | 26 | 3.3% | A | 67 | 8.5% | A | 67 | 2.8% |
| E | 96 | 12.1% | S | 64 | 8.1% | S | 64 | 2.7% |
| F | 14 | 1.8% | N | 58 | 7.3% | N | 58 | 2.4% |
| G | 21 | 2.7% | I | 45 | 5.7% | I | 45 | 1.9% |
| H | 31 | 3.9% | R | 41 | 5.2% | R | 41 | 1.7% |
| I | 45 | 5.7% | C | 36 | 4.5% | C | 36 | 1.5% |
| J | 0 | 0.0% | H | 31 | 3.9% | H | 31 | 1.3% |
| K | 6 | 0.8% | M | 29 | 3.7% | M | 29 | 3.7% |
| L | 23 | 2.9% | P | 27 | 3.4% | P | 27 | 3.4% |
| M | 29 | 3.7% | D | 26 | 3.3% | D | 26 | 3.3% |
| N | 58 | 7.3% | L | 23 | 2.9% | L | 23 | 2.9% |
| O | 71 | 9.0% | U | 22 | 2.8% | U | 22 | 2.8% |
| P | 27 | 3.4% | G | 21 | 2.7% | G | 21 | 2.7% |
| Q | 1 | 0.1% | Y | 16 | 2.0% | Y | 16 | 2.0% |
| R | 41 | 5.2% | F | 14 | 1.8% | F | 14 | 1.8% |
| S | 64 | 8.1% | B | 10 | 1.3% | B | 10 | 1.3% |
| T | 75 | 9.5% | K | 6 | 0.8% | K | 6 | 0.8% |
| U | 22 | 2.8% | V | 6 | 0.8% | V | 6 | 0.8% |
| V | 6 | 0.8% | W | 6 | 0.8% | W | 6 | 0.8% |
| W | 6 | 0.8% | Q | 1 | 0.1% | Q | 1 | 0.1% |
| X | 1 | 0.1% | X | 1 | 0.1% | X | 1 | 0.1% |
| Y | 16 | 2.0% | J | 0 | 0.0% | J | 0 | 0.0% |
| Z | 0 | 0.0% | Z | 0 | 0.0% | Z | 0 | 0.0% |

Table 2A.                Table 2B.                Table 2C.

Observations From Sub-Table 2A:

- Some letters do not appear at all!
- The letter 'e' appears significantly more often than the rest.
- The first 10 letters, have a significantly higher frequency than the other letters.
- Approximately, 6 letters appear very infrequently.
- The average frequency of a letter is 3.9%.

- The standard deviation from the mean frequency is 3.4%
- That any letter appearing with a frequency greater than the average plus the standard deviation (3.9% +3.4%=7.3%) is exceptional and unique. I.e. An identifiable letter.
- In the example above, 'E', 'T', 'O', 'A', 'S', & 'N' are exceptional and unique letters as we would expect.

Observations From Sub-Table 2C:

- Some letters do not appear at all!
- The letter 'e' appears as often as any other letter.
- The first 10 letters, with a significantly higher frequency from Sub-Table 2A, now appear as often as other letters.
- Approximately, 6 letters appear very infrequently.
- The average frequency of a letter is 2.1%.
- The standard deviation from the mean frequency is 1%
- Letters ['M', 'P' & 'D'] appearing with a frequency greater than the average plus the standard deviation (2.1%+1%=3.1%) are not exceptional and unique. I.e. An identifiable letter.

Table 3 below shows two sub-tables. Sub-table A on the left, shows the letters of the alphabet sorted by their original frequency prior to any attempts at frequency normalization. Sub-table B on the left, shows the letters of the alphabet; sorted by their new frequency if identities had been added to alter the frequency. We see previously frequent letters buried deep below. For instance, the frequent 'i', looks identical to an 'f'. The average 'm', now looks like the most frequent letter, an 'e'. We see that the frequent 'e' is now tied with 'o' and, the original top 10 letters [E,T,O,A,S,N,I,R,C,H], all have the same approximate frequency. In addition, remember to factor in, the additional 21 extra identity symbols. There are now, 10 original + 21 new symbols = 31 symbols: All with approximately the same frequency! The entire alphabet has 47 letters in total. (26 letters + 21 new symbols = 47 letters new alphabet)

[Each letter that has twin replacements is having its frequency divided by 3. So, there is the original symbol, plus 2 new identity symbols. (1+2=3). This is done for 9 of the first 10 letters. So, 9 x 2 = 18. The letter 'e', having such a high frequency, gets one more identity symbol. I.e. The frequency for 'e' is divided by 4 instead of 3. That requires 3 new symbols. So, 18 + 3 = 21 new symbols.]

| Frequency of Letters with New Symbol Set – Displayed in Original Frequency Order | | | Frequency of Letters with New Symbol Set – Displayed in New Frequency Order | | |
|---|---|---|---|---|---|
| E | 96 | 3.0% | M | 29 | 3.7% |
| T | 75 | 3.2% | P | 27 | 3.4% |
| O | 71 | 3.0% | D | 26 | 3.3% |
| A | 67 | 2.8% | T | 75 | 3.2% |
| S | 64 | 2.7% | E | 96 | 3.0% |
| N | 58 | 2.4% | O | 71 | 3.0% |
| I | 45 | 1.9% | L | 23 | 2.9% |
| R | 41 | 1.7% | A | 67 | 2.8% |
| C | 36 | 1.5% | U | 22 | 2.8% |
| H | 31 | 1.3% | S | 64 | 2.7% |
| M | 29 | 3.7% | G | 21 | 2.7% |
| P | 27 | 3.4% | N | 58 | 2.4% |
| D | 26 | 3.3% | Y | 16 | 2.0% |
| L | 23 | 2.9% | I | 45 | 1.9% |
| U | 22 | 2.8% | F | 14 | 1.8% |
| G | 21 | 2.7% | R | 41 | 1.7% |
| Y | 16 | 2.0% | C | 36 | 1.5% |
| F | 14 | 1.8% | H | 31 | 1.3% |
| B | 10 | 1.3% | B | 10 | 1.3% |
| K | 6 | 0.8% | K | 6 | 0.8% |
| V | 6 | 0.8% | V | 6 | 0.8% |
| W | 6 | 0.8% | W | 6 | 0.8% |
| Q | 1 | 0.1% | Q | 1 | 0.1% |
| X | 1 | 0.1% | X | 1 | 0.1% |
| J | 0 | 0.0% | J | 0 | 0.0% |
| Z | 0 | 0.0% | Z | 0 | 0.0% |

Table 3A.        Table 3B.

Table 4A.

| Letters & Identities Sorted by Frequency Prior to Normalization | | | |
|---|---|---|---|
| Original Letter or Identity | Tally | Normalized Tally of Letter & Identity | Frequency as a percent |
| E | 96 | 24 | 3.0% |
| é |  | 24 | 3.0% |
| â |  | 24 | 3.0% |
| ä |  | 24 | 3.0% |
| T | 75 | 25 | 3.2% |
| à |  | 25 | 3.2% |
| ü |  | 25 | 3.2% |
| O | 71 | 24 | 3.0% |
| Ç |  | 24 | 3.0% |
| ê |  | 23 | 2.9% |
| A | 67 | 22 | 2.8% |
| ë |  | 22 | 2.8% |
| è |  | 23 | 2.9% |
| S | 64 | 21 | 2.7% |
| Ï |  | 21 | 2.7% |
| Î |  | 22 | 2.8% |
| N | 58 | 19 | 2.4% |
| Ì |  | 19 | 2.4% |
| Ä |  | 20 | 2.5% |
| I | 45 | 15 | 1.9% |
| Å |  | 15 | 1.9% |
| É |  | 15 | 1.9% |
| R | 41 | 14 | 1.7% |
| Æ |  | 14 | 1.7% |
| Æ |  | 13 | 1.6% |
| C | 36 | 12 | 1.5% |
| Ô |  | 12 | 1.5% |
| Ö |  | 12 | 1.5% |
| H | 31 | 10 | 1.3% |
| Ò |  | 10 | 1.3% |
| Û |  | 11 | 1.4% |
| M | 29 | 29 | 3.7% |
| P | 27 | 27 | 3.4% |
| D | 26 | 26 | 3.3% |
| L | 23 | 23 | 2.9% |
| U | 22 | 22 | 2.8% |
| G | 21 | 21 | 2.7% |
| Y | 16 | 16 | 2.0% |
| F | 14 | 14 | 1.8% |
| B | 10 | 10 | 1.3% |
| K | 6 | 6 | 0.8% |
| V | 6 | 6 | 0.8% |
| W | 6 | 6 | 0.8% |
| Q | 1 | 1 | 0.1% |
| X | 1 | 1 | 0.1% |
| J | 0 | 0 | 0.0% |
| Z | 0 | 0 | 0.0% |

Table 4A. Continued.

Table 4B.

| Letters & Identities Sorted by Normalized Frequency | |
|---|---|
| Original Letter or Identity | Frequency as a percent |
| M | 3.7% |
| P | 3.4% |
| D | 3.3% |
| T | 3.2% |
| à | 3.2% |
| ü | 3.2% |
| E | 3.0% |
| é | 3.0% |
| â | 3.0% |
| ä | 3.0% |
| O | 3.0% |
| Ç | 3.0% |
| ê | 2.9% |
| è | 2.9% |
| L | 2.9% |
| A | 2.8% |
| ë | 2.8% |
| î | 2.8% |
| U | 2.8% |
| S | 2.7% |
| ï | 2.7% |
| G | 2.7% |
| Ä | 2.5% |
| N | 2.4% |
| ì | 2.4% |
| Y | 2.0% |
| I | 1.9% |
| Å | 1.9% |
| É | 1.9% |
| F | 1.8% |
| R | 1.7% |
| æ | 1.7% |
| Æ | 1.6% |
| C | 1.5% |
| ô | 1.5% |
| ö | 1.5% |
| û | 1.4% |
| H | 1.3% |
| ò | 1.3% |
| B | 1.3% |
| K | 0.8% |
| V | 0.8% |
| W | 0.8% |
| Q | 0.1% |
| X | 0.1% |
| J | 0.0% |
| Z | 0.0% |

Table 4B. Continued.

Tables 4A & 4B above shows the results of the frequency normalization process herein discussed. Table 4A lists the letters according to the frequency order prior to normalization. The identities of a letter are grouped with the letter itself. Table 4B shows the letters and their identities, *both*, sorted by their frequencies after normalization.

Observations: Several infrequently found letters rise to the top. Frequent letters are buried deep below.

Frequency analysis observations:

- There are 46 symbols
- The average frequency is 2.1%
- The standard deviation of the frequency of any given symbol is approximately 1% (0.995%)
- Frequencies between 1.1% -- 3.1% are within the standard deviation of the mean.
- 72% of the symbols (33/46) (approximately ¾) appear within a frequency range of the standard deviation (1%) from the mean (average).
- The remaining 18% of the symbols appear within a frequency range of 2% from the mean.

Comparison of Tables 4 with Tables 6:

- Compare the standard deviation of the expanded symbol set: 1% from Table 4; with the original standard deviation of 3.4% from Table 2B.
- All the letters from the expanded symbol set have a frequency less than the original average frequency. [Compare the highest frequency of the expanded symbol set: 3.7% from Table 4B; with the average frequency of 3.9% from Table 2B.]

With only a 1% difference between the majority of frequencies after normalization, the symbols are can not be differentiated after normalization if ciphered.

Also, the decipherer can not know, that extra symbols have been added. Even if an assumption (guess) that extra symbols have been added, one can not know which symbols are the extra symbols.

## 3.  ENCRYPTING THE BINARY
## 3.1  FUNDAMENTAL PRINCIPLES: BREAKING THE TRANSITIVITY AND CORRESPONDENCE OF ALPHABET TO BASE 2

[This section is not intended as a survey. It merely states all the computer science principles necessary to understand the cryptographic process.]

Encryption is an integral part of computing.

Electronic computers use an immense number of tiny electronic switches, measuring on and off, voltage positions of high or low. [MAL01] The status of these switches (on or off) is easily represented by a binary system – a system of only two possibilities. [MAL01] When stringed together, the many different combinations of just two possibilities, provide enough symbols for our needs. E.g. All the letters of the alphabet, letters of multiple alphabets, punctuation marks, numerals, different machine control codes, different machine operation codes, etc.

The string of switches are conventionally converted to the numerals 0 & 1 for easy representation. Also, these numerals (0 & 1) are conventionally grouped together. Due to the binary nature of the numbers and operation of current electronic computing machinery, a binary numeral system is used. [MAL01]

As binary numbers go beyond several digit places, they become unwieldy for human comprehension. An easy mathematical way, of representing large binary numbers, is to convert them into a number that represents an exponential multiple of 2. E.g. Base 16 (base$_{16}$), is the 4$^{th}$ exponential multiple of 2 (2x2x2x2=16). Using base$_{16}$ makes it easier for humans to conceptualize and deal with the large number of switches, their combinations, and the codes used for the combination of switches.

A base$_{16}$ number represents 4 binary digits. Two base$_{16}$ numbers are referred to as one byte and represents 8 switches. The possible number of combinations and permutations of all the binary switches (and digits) for 2 base$_{16}$ numbers is 256. [GOL001] This number (256) provides a sufficient amount of codes for all the symbols usually needed to represent linguistic data (letters and punctuation marks). [JEN01]

A code book to correlate the base$_{16}$ numbers to the language symbols is needed [JEN01]. There are two common code books. They are referred to as ASCII and EBCIDIC [COL01] [HOD01], with ASCII being more pervasive as it is used in PC computers.[2]

As a rule, most encryption systems take the ASCII value as a decimal number or; some numerical value standing for the language symbol, and scramble it somehow. (Either by encoding, enciphering or encryption, as will be explained in detail below.) [KAH11] Then, the new number is translated into a base 2 binary number.

If instead of using the standard base 2 binary representation for ASCII values, an alternate binary representation can be used. E.g. The letter "A" has an ASCII value of 65 in base$_{10}$. The number 65$_{10}$ in binary, base$_2$, is represented as 0100 0001. However, quite logically and mathematically, the same number, could be represented in a binary system – that is not base$_2$ – as 0001 1000 0000.

Indeed, as will be explained mathematically and cryptographically, there are many binary number systems: Some natural and many, unnatural.

The advantage to using alternate binary systems becomes clear when the decryption process is understood. While the decryption process will be explained below in detail, in very brief, it integrally involves a frequency analysis of the symbol set, the numbers representing the letters of the alphabet. Any given language will have an intrinsic frequency to certain letters and sounds. Some of which, will be high and, some will be low. By counting the frequency of symbols in a secret message and; matching those frequencies to the frequencies of the symbols within a given language, one slowly develops a correspondence between encrypted symbols and the alphabet. Thus, a message is decrypted. With more advanced encryption techniques, highly sophisticated mathematics are needed to determine the frequencies. But, the process remains the same: Find the most common letter, the 2$^{nd}$ most common letter, the 3$^{rd}$ most common letter, etc. [KAH05]

This decryption technique works, because, ultimately, a person is always working on only one symbol set. I.e. The letters of the known alphabet. The resultant encrypted letter (the output of the encryption) is always equal to a specific ASCII symbol. Meaning, however you encrypt your original text message, if the output is an "A", that "A" will always be an ASCII 65. And, any given ASCII number will always be equal to the same specific base$_2$ number. Serial ciphering will not alter the frequency of the letters in a message. No matter how many consecutive types of scrambling from alphabet, to base$_{10}$ to base$_2$ are used; the same inherent frequency of the source remains. [KAH11]

A basic transitivity exists:

$A_{\text{SOURCE SYMBOL}} \leftrightarrow B_{\text{ASCII BASE 16 NUMBER/CODE BOOK}} \leftrightarrow C_{\text{BASE 2 BINARY NUMBER}}$

So, no matter how many different substitutions you use to scramble an "A" – only the representation by number changes. You never change the codebook! The ASCII table always remains the same! And, you never change the base 2 binary number which is necessary to convert the ASCII number into a string of switches for the computer to work with!

If you break the transitivity; if the binary number is not a base$_2$ number; if the binary number is one of many different binary numbers; then, two totally different relationships have been scrambled. The $A_{\text{SOURCE SYMBOL}} \leftrightarrow B_{\text{ASCII BASE 16 NUMBER/CODE BOOK}}$ relationship is independently scrambled from the $B_{\text{ASCII BASE 16 NUMBER/CODE BOOK}} \leftrightarrow C_{\text{BASE 2 BINARY NUMBER}}$ relationship. In addition, the $B_{\text{ASCII BASE 16 NUMBER/CODE BOOK}} \leftrightarrow C_{\text{BASE 2 BINARY NUMBER}}$ is not a linguistic scrambling! Meaning, that there is no frequency analysis to be had, to figure out which decimal number is the most frequently encrypted, the 2$^{nd}$ most frequently encrypted, etc. By visual inspection, one can not know which binary system one is looking at.

(As will be explained below, there are some intricacies to specific binary systems that may exhibit or exclude certain visual patterns. But, this is not definite. And, a sufficient number of binary systems are available, that have no indicators whatsoever, to make scrambling of the binary system possible and logically irreversible.)

Thus, using the procedure above, a very strong encryption technique can be made with only the major drawback, that it is imperative to keep the keys or code tables as well as the binary system used, secret!

---

[2] While I could not find a citation or study to support this claim, it would appear, that as PCs are a ubiquitous commodity item and; mainframes a large ticket item reserved for government and industry; therefore, PCs are more prevalent than mainframes. And, as EBCDIC is an IBM code table, used for IBM mainframes [GAN01]; whereas ASCII is used on PCs; therefore: it appears that ASCII is more common than EBCDIC. In addition, ASCII is the backbone of Internet communications [HOD01].

To-date, neither the author nor any reviewer of this article has never read of any encoding or encryption device that uses this technique. Nor, has the author ever read of a proposal for using this technique. And, while there are mathematical works discussing binary systems and converting binary numbers to a standard form [KNO01] [WIK01]; the author has never heard or read of any one, applying such knowledge to encrypt a binary transmission. Usually, encryption is done on the letters, not on the binary representation.

## 3.2 DEFINITIONS:

[While it is customary to explain all technical terms in the beginning of a paper, prior to using them, doing so, in this instance, may give the impression of a survey. Therefore, this section has been made a glossary at the end of the article. The glossary defines all computer science, mathematical, linguistic and cryptographic terms used in this paper. The reader may wish to read the glossary before proceeding. Or; to continue reading and refer to the glossary for those terms which are unfamiliar.]

## 3.3 THE PROCESS:

The process is easily understood, but in application, may be more complicated:

1. Simply encipher or encrypt the letters of the message.
2. If a numerical value has already been assigned to each enciphered letter, then skip the next step.
3. If a numerical value has not been assigned to each enciphered letter, translate the enciphered letters into ASCII
4. Translate the ASCII into base 2 binary.
5. Cipher the base 2 binary with another binary system.

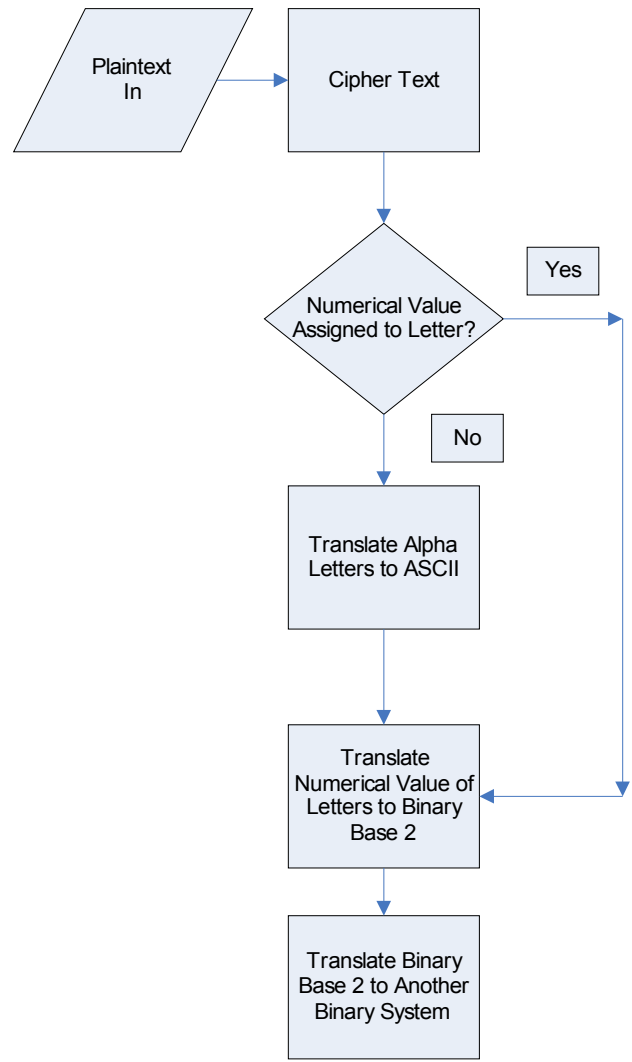

Flowchart A.

## 3.4 IMPLEMENTATION:

Now, we can get into the details of operation for encryption methods using alternate binary systems. From the definitions above, it is clear that aside from base 2, there are four natural binary systems: Fibonacci and Phinary, both standard and non-standard.

As for unnatural binary systems, many could be constructed. All that would be required is to either refer to a different series, other than Fibonacci. Or, exclude certain numbers from the natural order. Thus, by creating gaps in the number line, a new numbering system will be had. E.g. 0, 1, 3, 5 Or, in a binary format: "0000", "0010", "0100", "0101". If one is trying to encode, just the 26 letter alphabet, then, within a two byte space, 16 bits, many numerical systems could be had. (The significance of two bytes will soon become apparent.) In fact, for Fibonacci representation, only 5 bits would be needed for a reasonable minimum. Since encryption and confusion are the goal, there is no maximum to the number of bits we could use to generate artificial numbering systems. Quite a large number of binary systems could be generated with 16 bits alone. And, although it will take more bits to represent the message, the tradeoff will be more than worth the security of the data encryption, as will be explained below.

Using just 16 bits, to produce 256 symbols, which only requires 8 bits, will produce many extraneous, "don't care", symbols. The decryptor can not know which ones are the "don't' care" symbols. Nor does the bit sequence have to be consecutive. E.g. The first 2 bits could be part of the number, but the next 2 bits could not be part of the number. E.g. If instead of writing "1111 1111$_2$" for 256; one could write: "0000 1111 1111 0000$_{encrypted\ base}$". The inability of the computer to do math with such a binary number is irrelevant. The encryption is the only relevant matter.

The total number of 256 binary number encryption tables that could be constructed, for 256 numbers is, out of 16 binary digits (bits) : 1024*1023*1022*1021*…*768. [The first number can be one of any of the 1024 bit combinations. Now that one combination has been used, the next number can only be one of 1024-1, or 1023 bit combinations. Now that one more combination has been used, the next number can only be one of 1024-2, or 1022 bit combinations.] That is a mighty large number!

While it can be argued that this is just another cipher, and serial ciphering does not really add to the complexity of deciphering KAH11; the difference is, that neither the letters nor a numerical identity for the letters [A=1, B=2, etc.] are not being serially ciphered! The *binary numbers themselves* are being ciphered! And, it is not a mathematical formula to be deciphered! It's a code table!

Also, that the decipherers are expecting 8 bit groups to represent numbers. And, there is no way to know if this is a two 8 bit groups or one 16 bit group. It will be unknown as to how many different bit groups were used in enciphering a message if an 8 bit group does not work. And, if an 8 bit grouping does not work, it is unknown if that is because it is not the right bit grouping or; because of a complicated cipher/frequency confounding encryption scheme.

And, if just numbers are being encrypted, there is usually, no way to reverse that encryption. For example, I could encrypt all the numbers in a checkbook, with a numerical translation table. That is not decryptable. One can not reconstruct the proper binary numbers from encrypted binary numbers.

[Statistics are a funny thing. If it looks too good to be true, it is. There is something known as the "First Digit Phenomenon" and Benford's Law. It is a statistical law about distribution which explains a fact, that taking random numbers – first digits, such as in lists, usually produces a certain distribution of numerals. 30% for 1, 17.6% for 2, etc. This statistical law has been used in audits to find fraud. If it is used in the decryption of numerical data I do not know. The equation is (P=log(1+1/D). P – Probability, D – The Digit in question. E.g. For the numeral "1": P=log(1+1/1)=log(1+1)=log 2 = 0.30  Also, not all lists follow this law. So, even with Benford's law: The list in question may not be subject to Benford's law  [LIV003] [LIV004] Even if the list in question is subject to Benford's law, simply knowing something is wrong, does not tell what should be right.]

Each binary system will have its pros and cons in application – with computers or encryption.

As Huffman proved, the most compressed binary system is a base 2 system. [HUF01] A clear advantage to using base 2 binary. Which means, **conversely**, if base 2 is the *most compressed* binary system, then, *there must be other binary systems*!

### 3.4.1 Phinary

The Phinary system, especially the standardized Phinary system, uses many bits per number. It can be seen from simple inspection [See Table A, below, for the Phinary numbers 1 thru 10.], that Phinary numbers require lots of digits. The tradeoff in size will make the system produce very large bit-sized messages. Much larger than other binary systems. However, for small messages, this disadvantage may be of no significance. As often occurs, encryption is needed with short messages and not encyclopedias. So, in spite of the tradeoff, the increase in size and; subsequent transmission time, may be of no moment; considering the capacity of today's technology.

| Decimal | Base φ |
|---|---|
| 1 | 1 |
| 2 | 10.01 |
| 3 | 100.01 |
| 4 | 101.01 |
| 5 | 1000.1001 |
| 6 | 1010.0001 |
| 7 | 10000.0001 |
| 8 | 10001.0001 |
| 9 | 10010.0101 |
| 10 | 10100.0101 |

Table A.

### 3.4.2  *Fibonacci Representation*

While Fibonacci representation does not generate as many digits as the Phinary system, still, it requires more digits than a base 2 system. In fact, to express, the 256 characters of the entire ASCII table, will require 12 digits in Fibonacci. This is not that significant an increase in the number of bits used.

In addition, at the very minimum, four bits must be added to complete one byte to facilitate most computer operations. If four more bits—digits, extraneous digits are added to a 12 bit/digit Fibonacci number, then, the Fibonacci number appears just like two 8 bit base 2 numbers taking up 2 bytes.

Also, if four extraneous digits are added, then many additional bit patterns can be created and substituted for natural Fibonacci binary numbers. E.g. If the first four bits of every two bytes, is in a "don't care" state, then, the first four bits can be randomly filled with garbage data – noise. This will only add to the confusion of the binary number ciphering.

If every 12 bit Fibonacci number is padded with 4 bits, then, this is an increase of merely 1/3 the size of the entire message. Such a trade off in length is not a negative attribute given the current capacities of today's computers and transmission facilities. Confusion is paramount. A little extra space or time is of minimal concern.

In addition, if no padding of extra bits are used, but two 12 bit Fibonacci numbers are laid out, one after the other, then; two consecutive Fibonacci numbers appear as 3 two byte base 2 numbers.

To illustrate:

Using big endian, the highest natural Fibonacci number needed to express 256 is

"1111 0000 0000$_{FIB}$"

= $(1x122_{10}) + (1x68_{10}) + (1x44_{10}) + (1x21_{10}) + (0x13_{10}) + (0x8_{10}) + (0x5_{10}) + (0x3_{10}) + (0x2_{10}) + (0x1_{10}) + (0x1_{10}) + (0x0)$

= $255_{10}$.

Adding 4 extra bits, to fill out a byte, I could write the Fibonacci number "1111 0000 0000$_{FIB}$" as: "0000 1111 0000 0000$_{FIB}$" and express this number in two bytes. If I so desired, I could substitute, encipher, this Fibonacci number, "1111 0000 0000$_{FIB}$", with "1010 0000 0000 0000" or "1010 1111 0000 0000". Further confounding the encryption process and creating more binary systems.

So we see, that length, symbol boundaries and (byte) word boundaries are of significance in both encryption and decryption.

Since, if I were to employ enciphering of the binary system, as part of my encryption method, by picking and choosing from different binary systems; I could – as described above – create a binary system, made of 2 bytes, from which, I use only a sufficient set of symbols to express 256 out of the 1024 possibilities.

Indeed, using a two byte cipher for a one byte base 2 binary number, I could construct a cipher that would imitate a standardized Fibonacci number. This possible identity, demonstrates, that I could totally confound a message represented in a binary coding system by encrypting just the binary.

Furthermore: If I take a 40 character message, and transmit the same message as 4,000 bytes containing ciphered numbers, Fibonancci or not; unless the interceptor knows the length of the original message, there is no way to know, just how big a binary group might be and; how many binary groupings have been transmitted. Perhaps, forty 8 bit bytes were sent with a lot of garbage in between. Which means, that the binary encoding requires one hundred 8 bit bytes per character.

One could use base 2 binary, but, exclude all numbers that have an "11" sequence, in order to mimic a standardized Phinary binary system. Again, ambiguity provides obfuscation.

Practicality will limit the number of binary bases available for use. But even so, there are a sufficient number of possibilities to confound the process sufficiently to make *methodical deductive* decryption impossible.

### 3.4.3  Golden Sequence Representation

Use successive sequences of golden sequence symbols as numerals to represent numeric data for ciphering and encryption. Because of the order of the symbols, i.e. no symbol starts with a "0" or; that each symbol must be a combination of previous symbols; therefore, a string of golden sequence symbols can be broken up into individual parts.

Again, the symbols are purely binary. Again, there is no way of discerning these symbols from base 2 binary.

### 3.4.4  Base Prime Representation

One could define any number as a sum of prime numbers smaller than that number itself; with each prime number being used only once. Hence, if we use bits to represent the prime number sequence; e.g. 1, 2, 3, 5, 7, 11... In a fashion similar to Fibonacci representation, we could use prime number representation, to define each number.

For sure, we can conjure up other sequences as well, to use to mimic the idea of numeric representation, as we started with Fibonacci representation.

### 3.4.5  Boustrophedon:

If boustrophedon is applied to bit sequences, binary numbers, the result is NOT a mathematical inversion. (E.g. "0000 0001" becomes "1000 0000")  This is neither an additive, nor multiplicative inverse nor; is this multiplication by (–l) or some such procedure.  It is a physical inversion.  This is not decipherable by some mathematical calculation.  It is a pictorial encryption and the picture still looks legitimate.  There is no logical or mathematical way to know, what the original number was.

However, as in many forms of ciphering, even if the original symbols are swapped with new symbols; the original frequency is maintained.  Even if I do not know what the new symbols stand for.  And, so long as the language's frequency is maintained, it is decipherable. Or better put, translatable from the binary code to the original alpha letters.

But, if alternation (e.g. every other byte is inverted) or other variables are introduced (such as an encoding the letters with a key or an encryption method prior to inverting the binary); since one can not tell the difference between the pictures; the frequency analysis is confounded.  Once the frequency analysis is confounded, the message can no longer be decrypted. -- This will be true for any combination of methods that encrypt the binary numbers *and* confound the frequency analysis.

Alternating inversion of the bits with every other byte, would produce very interesting results.  Because, while it would halve the frequency of some letters, it would increase the frequency of other letters. Hence, the frequency distribution is disturbed.  In fact, by reason, it would halve the frequency of higher, more frequent letters.  Alternating boustrophedon would perform an incomplete frequency normalization.

In addition, there is no code table.  What is necessary for decoding, is the right sequence, starting position for inversion, and jump order (how many bytes to skip between inversions), etc.  These are parameters that are easily altered.

Certainly, if a complex mathematical formula was used to encrypt the data; and then, the resultant binary data was encrypted with boustrophedon; decryption would be impossible as correct mathematical calculations would be impossible and; deciphering numeric encryption is not possible.

However, we must remember, as is prone with encryption & encoding, espionage is engaged in, to steal the encryption algorithm or the codebook. (See [KAH14] for a good example of the necessity of stealing a codebook.) That would be true of encrypted binary systems too.  The only good – and sensible defense, is to continually change the encryption method or codebook.  [KAH07] [KAH08]  Encrypting the binary, especially with ciphered Fibonacci numbers, permits quick and constant alternate codebook generation.

From real life: Towards the end of WWII, the U.S. Army was changing codebooks for the U.S. forces in Europe at a rate of once every two weeks. [KAH01] The Japanese, who failed to change their codebooks, faced devastating results.  [KAH07] [KAH08]

For example, look at Table 5.  The left most column is a digit sequence.  The middle column is a base.  The right most column is the value of the sequence in base 10.  Every number is written as "11".  But, if I do not know what the base is, I do not know what the number means.  For all you know, it's "11" in base 256 or base 1024!

| Symbolic Representation | Base | Base 10 Equivalent |
|---|---|---|
| | | |
| 11 | Base 2 | 3 |
| 11 | Base 3 | 4 |
| 11 | Base 4 | 5 |
| 11 | Base 5 | 6 |
| 11 | Base 6 | 7 |
| 11 | Base 7 | 8 |
| 11 | Base 8 | 9 |
| 11 | Base 9 | 10 |
| 11 | Base 10 | 11 |
| 11 | Base 16 | 17 |

Table 5

The same would be true for expressing binary in a variety of different bases.

Take another example that has many significances. If I wanted to encrypt the numbers in a checkbook, and I use a simple cipher of adding one to a digit [9+1 becomes 0], then I transmit the numbers, that can not be decrypted. It is not possible. One needs some kind of mathematical reference, a total – correct or incorrect – to even know, if an encryption has been attempted.

Languages, as will be explained below in detail, have a natural frequency distribution of letters[3] [KAH15] – numbers do not! Unless there is a restriction on the possible numbers somehow, like map coordinates [KAH01] [VAU03] to clue one in somehow, there is no way of decrypting encrypted numbers. -- When this fact is factored into encrypting the binary, that one is encrypting numbers and not letters; then it becomes apparent that if the binary is encrypted, the binary can not be decrypted.

| A | 0.082 |
|---|---|
| B | 0.015 |
| C | 0.025 |
| D | 0.043 |
| E | 0.127 |
| F | 0.022 |
| G | 0.020 |
| H | 0.061 |
| I | 0.070 |
| J | 0.002 |
| K | 0.008 |
| L | 0.040 |
| M | 0.024 |
| N | 0.067 |
| O | 0.075 |
| P | 0.019 |
| Q | 0.001 |
| R | 0.060 |
| S | 0.063 |
| T | 0.091 |
| U | 0.028 |
| V | 0.010 |
| W | 0.023 |
| X | 0.001 |
| Y | 0.020 |
| Z | 0.001 |

Table B.
Frequency Distribution of Letters [VAU03]

This is very significant. Because often, secret messages often contain just numeric or monetary values. A very practical and historical example is that of agents representing buyers and bidders. The various agents, during the bidder process, must communicate with their home offices. But, the agents and buyers do not want their competition to know what price they are bidding.

To implement this, we need to remember that quite often, secret messages will have the minimum of information to get the message across. E.g. Go. Yes. No. Buy. Sell. Etc. Also, a code includes a prearranged agreement to symbols, which includes the symbol sequence. If an agent transmits just two numbers; the first the agent's bid, the second the competition's bid and, the buyer knows this sequence and; the numbers are encrypted and; that's all there is to the message; that is not a decryptable message.

---

[3] See Table B.

E.g. Simply add or subtract 5 from every digit to any sale price. Or, just add or subtract $5, from every sale price.

In sum, from the above examples, it becomes imminently clear, that enciphering the binary number has no connection with the encryption of the letters associated with the ASCII table.

Furthermore, if so desired, alternates to the natural binary systems can be employed increasing the number of possible ciphers for enciphering base 2.

Also, we must take into consideration, that while the ASCII table represents 256 characters; commonly, messages use far few characters. The rich character set of ASCII includes machine control codes, e.g. carriage return, new line, etc. The rich character set is not necessary for the data payload. Also, the ASCII character set includes graphics or other language letter symbols as well as capital and miniscule letters. Miniscule letters are often a redundancy that is eliminated in encryption.

In addition, the table could also be enciphered in and of itself. Adding to the confusion. E.g. The letter "A" could be represented by the number 32, instead of 65. This enciphering is also separate from the encryption of the letters of the original plaintext message. (But, this could be viewed as just serial enciphering, which does not really add to the strength of the encryption.)

## 4. SUMMARY OF BREAKING THE TRANSITIVITY AND CORRESPONDENCE OF ALPHABET TO BASE 2:

The ultimate goal of all language encryption is to confound the frequency analysis.

Every language has a frequency for the letters within the language. [KAH15] The frequency is different for each language. In English, the letters 'E', 'T', 'O', 'N', 'I', 'S' are the most common, with "E" being approximately 13% [KAH09] [KAH10] or 12.7% [VAU03]. This frequency does not really change. [KAH05] And, with more written samples, the more definite the frequency becomes. [KAH05]

The most basic kind of encryption, simple ciphering, replacing each letter with a different letter, will not alter the frequency of the letters. The cipher will look different from legible text, but, when tallied, the frequency of the letters will be the same. [KAH11]

E.g. "The quick brown fox jumped lazily over the sleepy dog." -- This sentence is used when teaching typing because it has every letter in the alphabet. An unnatural contrivance, true, which just strengthens the point that innate frequencies of letters exist. An analysis of the sentence reveals the following tabulation:

| Alphabetical Listing | | | Sorted by Frequency By Alphabetically | | |
|---|---|---|---|---|---|
| Letter | Tally | Frequency | Letter | Tally | Frequency |
| A | 1 | 1/44 = 2.3% | E | 6 | 6/44 = 13.6% |
| B | 1 | 1/44 = 2.3% | O | 4 | 4/44 = 9.0% |
| C | 1 | 1/44 = 2.3% | L | 3 | 3/44 = 6.8% |
| D | 2 | 2/44 = 2.5% | D | 2 | 2/44 = 2.5% |
| E | 6 | 6/44 = 13.6% | H | 2 | 2/44 = 2.5% |
| F | 1 | 1/44 = 2.3% | I | 2 | 2/44 = 2.5% |
| G | 1 | 1/44 = 2.3% | R | 2 | 2/44 = 2.5% |
| H | 2 | 2/44 = 2.5% | T | 2 | 2/44 = 2.5% |
| I | 2 | 2/44 = 2.5% | U | 2 | 2/44 = 2.5% |
| J | 1 | 1/44 = 2.3% | A | 1 | 1/44 = 2.3% |
| K | 1 | 1/44 = 2.3% | B | 1 | 1/44 = 2.3% |
| L | 3 | 3/44 = 6.8% | C | 1 | 1/44 = 2.3% |
| M | 1 | 1/44 = 2.3% | F | 1 | 1/44 = 2.3% |
| N | 1 | 1/44 = 2.3% | G | 1 | 1/44 = 2.3% |
| O | 4 | 4/44 = 9.0% | J | 1 | 1/44 = 2.3% |
| P | 1 | 1/44 = 2.3% | K | 1 | 1/44 = 2.3% |
| Q | 1 | 1/44 = 2.3% | M | 1 | 1/44 = 2.3% |
| R | 2 | 2/44 = 2.5% | N | 1 | 1/44 = 2.3% |
| S | 1 | 1/44 = 2.3% | P | 1 | 1/44 = 2.3% |
| T | 2 | 2/44 = 2.5% | Q | 1 | 1/44 = 2.3% |
| U | 2 | 2/44 = 2.5% | S | 1 | 1/44 = 2.3% |
| V | 1 | 1/44 = 2.3% | V | 1 | 1/44 = 2.3% |
| W | 1 | 1/44 = 2.3% | W | 1 | 1/44 = 2.3% |
| X | 1 | 1/44 = 2.3% | X | 1 | 1/44 = 2.3% |
| Y | 1 | 1/44 = 2.3% | Y | 1 | 1/44 = 2.3% |
| Z | 1 | 1/44 = 2.3% | Z | 1 | 1/44 = 2.3% |

Table 6.

Sub-Table A.          Sub-Table B.

It must be remembered, that this sentence is contrived to contain every letter in the alphabet. Even in a contrived sentence like this, the frequent occurrence of the letter "E" can not be avoided. In regular text, with a greater statistical sample (of letters), the frequency will be apparent.

Returning to the example sentence, notice, that if I replace all the 'e's with 'q's, how the frequency is unaltered: "Thq euick brown fox jumpqd lazily ovqr thq slqqpy dog." -- There is now one 'e' and 6 'q's. The letters look different, but the frequency is the same.

Serial enciphering, using one cipher substitution after another, may appear complicated, but in reality, it does nothing to alter the difficulty of decryption. Because, once again, the frequency analysis is unaltered. And, it is entirely a binary correspondence.

E.g. If I repeat the substitution from the above example, and now replace all the 'q's with 'x's; then I have: "Thx euick brown foq jumpxd lazily ovxr thx slxxpy dog." There are now 6 'x's, representing the same letter. And, only one 'q', representing the only 'x' in the sentence.

What more complicated ciphering does, is to alternate, which letters replace other letters. Often, this is done with a key. So, that given a word, like "sleepy", the replacement for the 1$^{st}$ 'e' will be 'q', but the replacement for the 2$^{nd}$ 'e' will be 'x'. Now, when encoded, we have "slqxpy". Now, the frequency of the appearance of the letter 'e', has been changed.

As interesting as the subject is, I will not go into the complexity of the keys and methodology of decryption. (I refer interested readers to David Kahn's excellent book, The Codebreakers. [KAH01]) Suffice to say, some trace of cyclic repetition remains and can be ultimately factored out. Albeit by difficult mathematical means and computers, but any key, no matter how long, so long as it repeats, can eventually be decrypted. This is why large prime numbers are so important to encryption. Because one can go on for millions of digits, without a repetition ever occurring. Think of digits as characters, or alterations to characters.

As David Kahn, a great historian of cryptography noted [KAH05] [KAH11], people often mistakenly think that it is the complexity of the encryption that makes something decipherable. But, this is a fallacy. Indeed, the method herein proposed, is a simple, and yet, quite elegant way, of encrypting and making something undecipherable.

But, by employing alternate binary systems, and encoding the binary translation of an enciphered message, one is not adding to the same target of substitution. The binary encoding does not just add another layer of substitution to the *letters*, albeit a bit more complicated. Encrypting the binary number system is a different target of encryption than the letters of the message.

Therefore, the ternary correspondence of letter to ASCII to binary is irrevocably altered.

Disregarding the ASCII table. One can conceive of the encryption used with computers, as the substitution of a base 10 (decimal) number for a letter; which is then converted from base 10 to base 2 binary. Even this simple correspondence of base 10 to base 2 binary is broken with binary encryption.

[In the examples to follow, { }, ( ) & [ ] are used to pair corresponding symbols and connote a relationship. Symbols outside an enclosure are enciphered symbols referencing the symbols inside the enclosure marks.]

Conventional encryption only encrypts the letters of the original plaintext message. There is a simple logic, that $A_{SOURCE\ SYMBOL}$ = $B_{ASCII\ BASE\ 16\ NUMBER/CODE\ BOOK}$ = $C_{BASE\ 2\ BINARY\ NUMBER}$. A=B=C. And, no matter what kind of cosmetic alteration to "A" is done, it remains, that A=B=C. Only one variable, "A", is encrypted. But, the relation, A=B=C is still valid. E.g. In binary base 2, using ASCII: "A" = 65 = "0001 0001". If some letter, "Q" for example, is converted to another letter, "A" for example, by some encryption technique, the correspondence has not been changed. E.g. "Q" = [(65 = "0001 000") = "A"] So, really, only one symbol has been encrypted. That limitation (of only one symbol having been encrypted) makes decryption possible.

However, if an alternate binary system is used, the circumstances are different. E.g. "A" = 65 = "0001 1000 0000$_{FIB}$" already alters the correspondence, A=B=C by only one enciphering. [("A" = 65) = "0001 0001$_2$"] = "0001 1000 0000$_{FIB}$"). Such an alteration would be an enciphering "0001 1000 0000$_{FIB}$" = "0001 0001$_2$". If an additional enciphering is used, e.g. "A" is exchanged with "Q", then the transitivity is broken. E.g. {"Q" = ("A"} = {65) = ("0001 0001$_2$") = "0001 1000 0000$_{FIB}$"). A≠B≠C. And; enciphering $B_{ASCII\ BASE\ 16\ NUMBER/CODE\ BOOK}$, would only complicate matters more, because any frame of reference is now gone. E.g. {"Q" = ("A"} = {[65=54]) = ("0011 0110$_2$"} = "0010 0100 1000$_{FIB}$").

[It is more than serial enciphering, because the target of enciphering is different; the binary system is not directly attached, referencing, the original plaintext. Therefore, when the binary system is enciphered, it is not an additional enciphering of the original plaintext.]

[In addition, enciphering the binary is not a mathematical change, but a pictorial change.]

The frequency analysis is confounded because you don't know what you are counting. Meaning, let's assume the original message is in unencrypted English – Plaintext. You have a bit stream that was enciphered with "a" binary system. You have no way of knowing which binary system it is. You do not know if you should tally different individual bytes or; every different set of two bytes; or every 12 bits. The tally will not necessarily generate a frequency analysis similar to English. None of the frequency analyses generated need be similar to English. More than one frequency analysis may be similar to English. The message may be too short to confirm a frequency analysis. If a sophisticated encryption algorithm was first applied to the letters, to substitute for other letters, to severely confound the frequency analysis, there may be no frequency observable. If some unnatural binary coding system was used; with or without a sophisticated encryption algorithm, a frequency analysis will not be apparent. Best case scenario, you have to engage in many frequency analyses, of several different bit lengths.

## 5. DECRYPTING:

### 5.1 Normalized Letter Frequency

This would require assuming the language of the ciphertext and its corresponding frequency. Then, guesses would have to be made to reconstruct the symbol set (of the alphabet). Apply the symbol set guessed. See if the decrypt makes sense. The magnitude of the decryption process would be measured in the factorials of the possible combination of symbols.

Spacing and punctuation—which may or may not have been included—would have to be taken into account.

It must be noted that the combinations of the frequencies will not be exact. Frequencies are probabilities. Tolerance factors will have to be introduced to guessing symbol sets. [The frequency of the letter 'e' may be 11%, 12%, 13% or even 14%. Any combination of symbols within the range must be considered.] I have never seen an actual letter frequency which is a whole number.

Also, frequencies vary with text and probably with context.

Also, the smaller the text, the greater the possibility the frequency will deviate from the standard frequency.

One option, is to select an assumed subset of symbols to be equal to the most frequent letter, 'e', then; 't', etc. Then apply the assumed reconstructed symbol set to the ciphertext and see if it makes sense.

However, if any kind of encryption was done to the ciphertext after replacing the ordinary alphabet with the new symbol set then; the regroupings and tests will be on the wrong symbols and combinations of symbols. There is no way to know what the original symbols were and which encrypted symbols refer to which symbols in the (new, revised) extended symbol set.

This is not decryptable by method—uncrackable—unbreakable. Brute force is useless.

### 5.2 Encrypted Binary

As the encryption possibilities for the binary system increase, so does the decryption possibilities decrease.

Again, look at Table 5. If one does not know the value of the sequence of digits "11", one does not know the base. I can not. I can only guess or assume – which is a guess.

Sometimes, when it is known, that the numbers must have certain values, decryption is possible. This scenario occurs in military applications when determining encrypted map coordinates. [KAH01]

Again, if I encrypt all the numbers in a checkbook with some cipher key, that can not be decrypted. A control number might indicate inaccuracies, but a control number will not indicate a deciphering method. Certain arithmetic manipulations might indicate transpositions (exchanging certain numbers for others) within a given sum. But, a cipher will not be found.

Again, Benford's law might indicate something is amiss. But, Benford's law will not tell you what the correct numbers should be.

The key to decryption of text expressed as numbers is in the relationship of the numbers to the letters. When numbers represent letters, decryption is commonly done by frequency analysis of the numerals representing the letters. [KAH05]

When employing frequency normalization, there is no frequency analysis to be had. Hence, the standard and common method of decryption innately fails.

As explained in a previous paper [ZIR01], if the frequency is altered, then, decryption becomes difficult. If the frequency is sufficiently altered, then decryption is impossible.

What an encryption key does is to alter the frequency of the appearance of letters, especially, even in any cyclic fashion of any kind. With longer and longer keys, one simply lowers the frequency more and more. When one uses a large enough prime number as a key to encrypt a message; essentially, what one has done is, to alter the frequency of appearance of each symbol to "1".

To undo frequency normalization, theoretically, one could guess and tally up different symbols, in different permutations, assuming a valid, normal frequency distribution. And; attempt to decipher the message based upon these guesses. However, if you compare Tables 1A & 2A; you will see that there is a difference between the frequency for the letter 'e', the most common letter in the alphabet between the two texts. Estimated frequencies are not exact. It is only in theory, that some such brute force method may be able to decipher such an encryption. That theory requires many variables to become known constants. This is not reality. There are too many variables and unknowns. In practicality, I do not think decryption of frequency normalization is plausible.

For Example: Using a binary encryption scheme without adding new symbols, -- assuming you know the source language, assuming there was no substitution or scrambling of the letters and; assuming no extra dummy letters have been added – at the very minimum – from the methods listed in this paper alone – several frequency analyses have to be done. After all, it could be a 16 bit binary system. Or; an 8 bit binary system, but each 16 bits is two 8 bit symbols. Or;, a 12 bit binary system, that requires two bytes to represent itself, so it appears as two 8 bit symbols. And, if a larger binary representation, such as φ is used, the number of possibilities and attempts increase. Also, one has to account for big endian, little endian possibilities?

In fact, one could use base 2 binary, but skip each number that contains a "11" in order to mimic a standardized Phinary binary system. There is no way to discern the difference.

If it is only several possibilities as described above, then, there are several analyses to review. Some will be gibberish, and one will be valid. Assuming, nothing was done to confound the frequency. But, extra letters could have been added in the original text. Or, a 12 bit binary system, could have been coded with extra dummy numbers that are not valid, to distort the frequency distribution. In fact, it is conceivable, that sufficient leeway is possible, to remove the frequency distribution altogether with dummy letters!

While one might argue, that if a Fibonacci numbering system is being used, then, the (unique) properties of the Fibonacci system will be discoverable. Then, the message could be decrypted. – This is a fallacy. I will explain:

1. The Fibonacci numbers are not being written – in or out of sequence.
2. **_References_** to Fibonacci numbers are being written – NOT Fibonacci numbers!
3. The Fibonacci numbers are only *correlated* to a bit pattern.
4. But, the bit pattern itself does not express a Fibonacci number!

Using a bit pattern correlating to Fibonacci numbers to represent positive integers results in non-Fibonacci numbers. -- I.e. The Fibonacci sequence has the series of numbers, starting from 0 or 1, that satisfy the formula, $f_n=(f_{n-1})+(f_{n-2})$. The series is {0,1,1,2,3,5,8,13…}. While the numbers 2 & 5 are Fibonacci numbers; their sum 7 is not. Using a bit pattern (little endian) "0010 1000$_{FIB}$" that references the Fibonacci series to indicate the addition of 2+5; does not produce a Fibonacci number. [Even in base 2, the bit pattern "0010 1000$_2$" = $33_{10}$. Thirty-three is not a Fibonacci number either.]

5. The unusual properties of the Fibonacci sequence only appear, when the Fibonacci numbers are used in a way that takes advantage of the numbers' order within the Fibonacci series. I.e. (fn=Fibonacci number) fn, fn+1, fn+2, fn+3….

Example:

$$[(f_n+1)/f_n]=[(f_n+2)/(f_n+1)]=[(f_n+3)/(f_n+2)]=[(f_n+4)/(f_n+3)]… = \text{The Golden Mean [LIV02]}$$

Using a bit pattern correlating to the Fibonacci numbers to represent letters, will not produce Fibonacci numbers in sequence. Hence, you can not test sequential numbers. If you can not test sequential numbers, the tests will fail.

It should be intuitive that any bit pattern could be correlated to a subset of the Fibonacci sequence or; any other sequence for that matter. There is no way to determine what sequence, if any, the bit pattern is referring to.

Decryption techniques other than frequency analysis are now necessary; such as capturing known cleartext messages and their corresponding ciphertext. It will become immediately apparent, that there are more bytes than a one to one correspondence would require. That could mean many things. But, it is a clear indication that a pure 8 bit/byte binary representation is NOT being used! Possibly, every other byte is a dummy byte. Or, some other algorithm may be used to insert dummy bytes. Someone may decide to encrypt each character with more than 2 bytes. It is unknown how many bits did used for a valid character.

Of course, the more bits used, the longer the message, the longer the time involved in transmitting that message. But, someone may decide, the security is worth it.

## 5.3   Brute Force

Brute force techniques are predicated upon the assumption, that the method of decryption is known and; there are a finite – a very large number – but finite, number of possible solutions. It is just the amount of time to test each and every possible solution that is the obstacle.

For example, dictionary attacks for passwords. A password is known to exist and *all* the possibilities of the password are known. Exactly which characters can make up the password are known. The possible lengths of the password are known. All the possible permutations and combinations of characters are known. There is a limited set of alphanumeric possibilities to the password. A very large set, but a limited and known set. A person trying each and every possible password, one at a time, will take too long to penetrate the security. But, a computer could do so in a half hour or, overnight. Or, with a distributed network, over the course of months or years.

Brute force attacks for longer messages and more complicated encryption methods usually attack targets assuming known methods of encryption were used, especially methods that use keys for ciphering. The long keys, the complicated mathematical equations, are all known. Computing the questions or testing the keys takes a long time. But, it is doable. The time obstacle is diminished by distributed computing or supercomputer power.

But, the basic principle of brute force decryption is:
1. The methods of encryption are known.
2. All possible keys are known – even if it is a very large number.
3. The numerical representation of the letter is known and; if reversing the mathematical encryption is necessary, it is possible to do the math.
4. It is just a matter of processing time to perform all the calculations.
5. The frequency of the letters in the ciphertext has been maintained – in some form.

When applying the rules necessary for a brute force decryption to the encryption methods described in this paper:
1. The methods of encryption are known. – We just discussed them.
2. While the implementations of the methods of encryption, that we discussed, are finite; the possibilities are infinite: Unlike the possible number of passwords to a given system, which is finite.
3. As for keys:
    a. While keys may be used and are known, they are not necessarily an integral part of the process. But, if keys are used, they are known.
    b. Since the encryption method is based upon pictorial representations; using combinations of 1s & 0s, in different or same length strings; to reference subsets of numbers: The question is, are the combinations of 1s & 0s, in different or same length strings finite and/or the number of subsets of numbers these representations refer to; finite? – Since I can always add one more 0 or 1 to the string, these pictorial representations are infinite. As for the number of subsets of numbers referred to, no matter how large, it is finite.
4. The numerical representation of the letter is **not** known and; if reversing the mathematical encryption is necessary, it is **impossible** to do the math until the numerical representation is known.
5. The frequency of the letters in the ciphertext has **not** been maintained! -- A crucial difference!

### 5.3.1 Brute Force Fails When Applied to an Infinite Number of Possibilities

Once an element of infinity has been introduced (point 2 & 3b), brute force fails – as a method. Brute force may provide a lucky guess. But, brute force will not definitely provide an answer. This is fundamental difference in applying brute force to binary encryption as opposed to key based encryption.

Example: Currently, with a dictionary type attack on a 5 letter password, the total number of possible passwords are $256^5$. A big number. Not humanly possible, unless one is dedicating one's life to the solution. One can think of medieval mathematicians calculating the values of sines, cosines and logarithms. For a computer: It's just a half hour's work. It's not a guess! It's an algorithm based upon permutations.

### 5.3.2 Brute Force is Inaccurate When Applied to Frequency Normalization

Against frequency normalization alone, not in conjunction with any other encryption method, brute force may, by assembling all the possible permutations, reconstruct the original sequence.

Consider: Pasting together symbols with different frequencies, to ascertain which sets of symbols represent actual alphabetic frequencies, may follow a method. However, the frequencies of letters that we use, are only theoretical – not actual. The actual frequencies fluctuate and differ between real messages. Compare Tables 4B & 6. Even for high ranking letters, the frequencies of appearance are almost all not the same. – The frequencies of appearance may be similar for the appearance of the same letter in different texts; but usually the frequency of appearance is not the exact same frequency in two separate texts. [E.g. In comparing Tables 6 & 2B, the letter "E" appears 13.6% in Table 6 vs. 12.1% in Table 2B.] So, even though we have a method, it is inaccurate and we may not succeed.

### 5.3.3 The Possible Number of Binary Representations are Infinite

With the binary representation encryption methods described in this paper, there is no algorithm to decryption. We start by guessing one method, then another, then another. What if the encryption uses a different implementation than one discussed in this paper? We have no method to try all different implementations.

## 5.3.4 Alternate Bit Patterns Appear Similar and Can Not Be Differentiated

Essentially, by binary representation redundancy, we have introduced a parameter into the encryption method that can not be discerned by a computer. Akin to CAPTCHA, but the lack of identity applies to humans as well. The computer can not tell by looking at or, inspecting, the bit patterns; which binary representation/encoding was used. Because, all possible bit patterns, are valid bit patterns, for many different possible representations. E.g. If a sigma is used in English writing, something is wrong. A sigma is not an English letter. But, a Fibonacci or prime number binary representation, is a valid bit pattern for base 2.

Also, as the numeric representation is not known:
1. We have no way of mathematically solving for equations that may have encrypted the data.
2. The issue of reconstructing symbol subsets arises again. Only this time, for the numbers themselves. This encryption does not have a frequency analysis to use as a basis for reconstruction. We have no method of reconstructing the numerical references, if frequency normalization was used on the numerals.

   [Given the example above in Table 4, if using the Fibonacci representation for the numbers 1 through 46, many numbers can be represented by different sums of Fibonacci numbers. This encrypts the numerals. – Not the letters. This complicates reconstructing the symbol sets with the frequencies of letters and impedes such reconstruction.]

One can argue, that, if known binary numbers, were used for these symbols, you could algorithmically – by method – go through a large number of permutations and reconstructions to attempt to guess the correct correspondence of sets of identities to letters. Although, this will require a lot of computing power and time.

Counterpoint: While there may be an algorithm that can give every possible permutation to reconstruct the correct frequencies from a frequency normalization; there is no algorithm to determine the binary number encryption. So, if you do not know what you are counting, how can you reconstruct the frequency?

Perhaps brute force could be used against frequency normalization, if there was no binary number encryption. Then, by adding symbol frequencies together, to create a table that matches the normal frequency distribution, one could try to recreate the message. Yes, this would involve many permutations until a correct table would be made. Which is par for brute force. However, it will require human intervention, CAPTCHA, to inspect each possibility for correctness.

However, in combination with binary number encryption, decryption is not possible. Because, you do not know what symbols to count. You can not create a tally for frequency analysis or reconstruction.

Brute force fails against a theoretically non-decryptable encryption method such as one time key encryption. Likewise, I maintain that brute force fails when the two methods together: encrypting the binary numbers in another base besides base 2; along with the combination of letter frequency normalization are used together. Because, the combination of the methods is theoretically not decryptable.

In sum, just as one time key encryption is theoretically proven to be non-decryptable; so to the *combination* of ciphering the cleartext *and* encrypting the binary numbers separately, results in a lack of correlation making decryption impossible. Likewise, frequency normalization, especially when coupled with encrypting the binary numbers separately, result in a lack of frequency making decryption impossible.

## 6. SUMMARY

Several encryption methods are proposed:

1. An encryption method that targets the binary numbering system alone. This method uses other binary numbering systems, both natural binary systems such as Phinary and Fibonacci, as well as unnatural binary systems, to replace the base 2 system.

2. A second encryption method of ciphering the text *and* encrypting the binary numbers. This method provides a theoretically undecipherable system.

3. A third encryption method of frequency normalization; using a sufficient number of identity letters for high frequency letters. This reduces the frequencies of high identity letters and introduces additional letters into the alphabet. This method is a strong encryption method, if not decipherable. This method may be decryptable by brute force.

4. A fourth encryption method of frequency normalization *and* encrypting the binary numbers. This method also provides a theoretically undecipherable system.

Process #1:

1. Translate the letters of the plaintext to ASCII values.
2. Cipher the base 2 binary values of the ASCII values to another binary base.

Flowchart of Process #1:

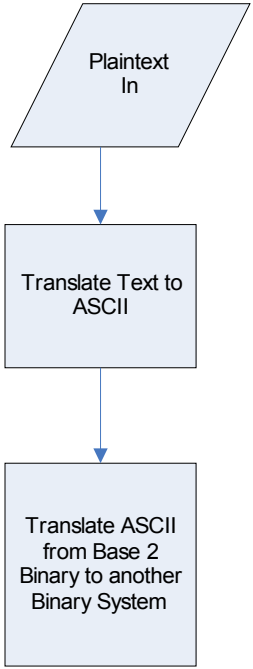

Flowchart B – Process #1.

Process #2:

1. Cipher the letters of the plaintext.
2. Translate the letters of the plaintext to ASCII values.
3. Cipher the base 2 binary values of the ASCII values to another binary base.

Flowchart of Process #2:

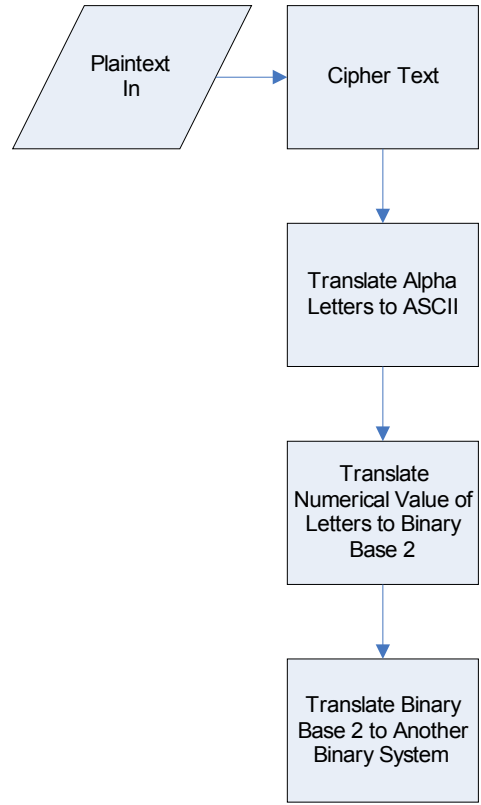

Flowchart C – Process #2.

Process #3:

1. Add 3 symbols as identities for the letter 'e'.
2. Randomly replace the letter 'e' with identities.
3. Add 2 symbols as identities for the 9 next, most frequent letters in the alphabet.
4. Randomly replace the nine most frequent letters with their identities.

Flowchart of Process #3:

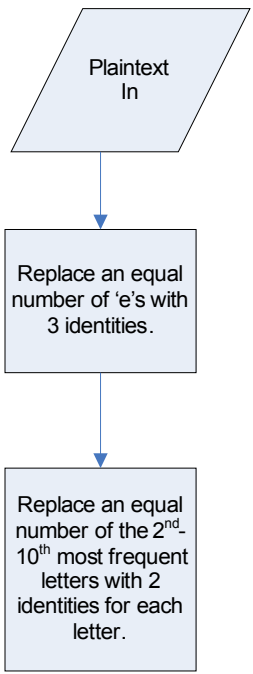

Flowchart D – Process #3.

Alternate implementations could include using a formula to figure out how many identities to add, per letter. This would be done by calculating the average frequency. Then, dividing any given frequency by the average frequency, to ascertain the number of identities necessary to generate, for any given number.

Process #4:

1. Add 3 symbols as identities for the letter 'e'.
2. Randomly replace the letter 'e' with identities.
3. Add 2 symbols as identities for the 9 next, most frequent letters in the alphabet.
4. Randomly replace the nine most frequent letters with their identities.
5. Cipher the text.
6. Encrypt the binary.

Flowchart of Process #4:

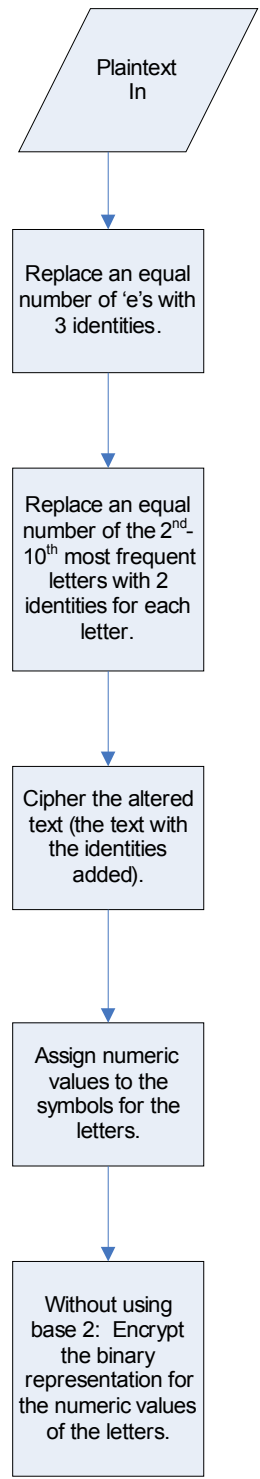

Flowchart E – Process #4.

Alternate implementations could include using a formula to figure out how many identities to add, per letter. This would be done by calculating the average frequency. Then, dividing any given frequency by the average frequency, to ascertain the number of identities necessary to generate, for any given number.

## 7. INFINITE COMBINATIONS

I hesitated from saying that these methods of encryption were not decipherable. Not because they were not. But, because they have an attribute in common with one time key encryption; in that the encryption is not decipherable by method; however, they are decipherable by brute force. The reason one time key encryption is decipherable by brute force is; that the number of possible keys is finite. However, I have now reasoned how to make the methods proposed, have an infinite number of possible combinations. Thereby, defeating brute force decryption as a method of decryption.

Once upon a time, before the computing power of our day, one time key encryption could not be guessed. However, today, with current computing power, a brute force *method* will decipher one time key encryption. Because, the number of keys are finite. Albeit, a very large number of possible keys. But, still, a finite set. So, each possible key is individually tested. If however, one could make an *infinite* set of keys, or permutations, one could make an encrypted message theoretically non-decipherable.

While I had the beginnings of such an idea, it was not fully developed. Now, I have come upon a way of making the above methods applicable in an <u>infinite</u> number of different ways.

One possibility: Using the method of altering the picture, by bit-slicing, meaning: combing the plaintext bit pattern, with the golden sequence, one can produce ciphertext. As the golden sequence is self expanding, the golden sequence can be enlarged to be as large as necessary for any given message.

For example: Using some known algorithm, add extra bits. E.g. After every third bit, insert a bit from the golden sequence. The ciphertext produced, similar to the ciphertext produced by the method of reversing the bits, has an altered pattern. The alteration is pictorial and not based upon math per se. The target of the ciphering is the ASCII. The result of the ciphering is indistinguishable from ASCII. The process is not reversible by method. There is no way to know if the original ASCII has been altered. And, there are an infinite number of possible ASCII cipherings.

Ex. The merging of plaintext with the golden sequence on the third bit of each byte:

ABC -- Plaintext

65 66 67 -- ASCII

0010 0001 || 0010 0010 || 0010 0011  -- Binary

001-0 00-01 || 0-010 –001-0 || 00-10 0-011 – Divided in Triplets

1011010110110  --  Golden Sequence

0011 0000 || 0101 0101 || 0010 0001 || 1000 0111 -- Merged

48 85 65 135 – Ciphertext ASCII

0 U A Graphic -- Ciphertext

Also, it was noted, that the increase in size of the ciphertext over the plaintext, is acceptable, in order to achieve security. This gives an infinite number of possible combinations. Which, is much better than a finite number of combinations: The drawback of any key encryption method. Because, all key encryption methods have a minimum and maximum size. One size is the case when the minimum equals the maximum. Also, all key encryption systems have a finite set of symbols (numbers) from which to select for each digit/position/character. So, the possible permutations and combinations to create keys is finite.

One is not limited to the golden sequence. There are an infinite number of sequences, irrationals, transcendentals, etc. to base such a method on.

Also, what is merged can be mixed up, in an infinite number of ways. The starting point in the irrational or golden sequence can be altered. Non-sequential digits can be used for merging. Not every byte has to be merged. Etc.

Ex. Sin 49º 50' 39"; every third digit of Sin 49º 50' 39"; Sin 49º 50' 39" spliced into a round robin of every set of $1^{st}$, $2^{nd}$, $3^{rd}$ and $4^{th}$ bit position.

One can go on and on with, more and more possible combinations. The point is, the possible number of combinations is infinite: Practically, as well as theoretically. This is because one can always add another position or byte. In addition, each new combination will be unique, with its own unique frequency analysis; which does not correlate to the original plaintext.

Now, that an infinite number of possibilities has been introduced, deciphering, by method is truly not possible. While guessing may work, there is no method.

Brute force guessing would require scanning for familiar bit sequences. Also, if the algorithm is known, a simple XOR will decrypt the ciphertext. Also, the security as well as transmission of algorithm choice and merging sequence, is no better than keyed encryption. But, I contend, that there are an infinite number of *"<u>unfamiliar</u>"* irrational sequences to choose from to make guessing impractical. I maintain that while guessing is possible, it is highly improbable.[4]

---

[4] Credit goes to Dr. Gertrude Levine (Fairleigh Dickinson University) for her criticism and disagreement, which helped me coalesce these thoughts.

## 8. FURTHER RESEARCH

[Original content deleted.]

The ensuing discussions about the original article has led me to understand that there are presumptions about one time key pad and its implementation. These assumptions lead to building in vulnerability and susceptibility to brute force attack. Other implementations would eliminate that vulnerability. That is out of the scope of this article.

## 9. CONCLUSION

Essential to achieving the goal of this project, was to use a different perspective. Hitherto, throughout the history of cryptography; all enciphering and encryption methods sought uniqueness to obfuscate the data – unique encoding of each letter in a message, unique keys. This method uses the opposite approach: ambiguity – multiple letters for the same letter – to obfuscate the data.

In addition, this work demonstrates the aphorism, that a mathematical proof should be like a poem. The methodology is simple and requires few steps as well as little effort.

**Author's Bio:** Givon Zirkind received his Bachelor's in Computer Science from Touro College and; his Master's in Computer Science from Fairleigh Dickinson University, both schools are located in the USA. His career has involved computer operations; software engineering; design and management of business applications with extensive database programming and management; Internet, web page design and implementation; e-commerce solutions, Google analytics; computer communications, data transfers and telecommunications; data conversion projects; reverse engineering of data and legacy software; being a published author and editor of a technical journal; teaching and; automated office support. His research work includes AFIS data compression and independent genetic database development and research. He may be reached at his email: GIVONZ@HOTMAIL.COM.

## 10. ACKNOWLEDGMENTS

To my grandfather for all his support in all my endeavors.

Allen Scott Gerner, colleague and friend, for his assistance with compiler operations, access to his library and encouragement. B.S. Computer Information Science, NJ Institute of Technology; M.S. Computer Science, NJ Institute of Technology.

To Ramona Brandt for donating resources and support to this project.

Dr. Larry T. Ray, R.I.P., Ph.D. Mathematics/Computer Science, Stevens Institute of Technology (NJ), formerly professor of computer science, Fairleigh Dickinson University, for his mathematical evaluation and support in this project.

To Jack Lloyd and all those on his Cryptography List, randombits.net for all their input and comments. Travis H. Jeremy Stanley, and others.

## 11. REFERENCES

[BAM01] Bamford, James; The Puzzle Palace: Inside America's Most Secret Intelligence Organization, Penguin Books, USA, 1983, ISBN: 0-14-00-6748-5

[BET01] Understanding Big and Little Endian Byte Order, http://betterexplained.com/articles/understanding-big-and-little-endian-byte-order

[BLA01] Blanc, Bertrand; Maaraoui, Bob; Endianness or Where is Byte 0, http://3bc.bertrand-blanc.com/endianness05.pdf

[CAP01] CAPTCHA: Telling Humans and Computers Apart Automatically; http://www.captcha.net/

[COL01] Collin, S.M.H.; Dictionary of Computing, Fourth Edition, Peter Collin Publishing, 2002, ISBN: 9781901659467; See entries for ASCII and EBCDIC

[DAV01] DaVinci, Leonardo; The notebooks of Leonardo DaVinci; Konecky & Konecky, Old Saybrook, CT; ISBN: 156852448X, Translated by Edward MacCurdy

[EUC01] Euclid, translated by Sir Thomas L. Heath, Book 6, Dover Books, USA, ISBN: 0-486-60089-0; Proposition 30

[EUC02] Euclid, translated by Sir Thomas L. Heath, Book 2, Dover Books, USA, ISBN: 0-486-60088-2, Proposition 11

[FRI001] Friedrich, Johannes; Extinct Languages, Dorset Press, New York, 1989, ISBN: 0880293381

[GAN01] Ganssle, Jack and Barr, Michael; Embedded Systems Dictionary; CMP Books, 2003, ISBN: 978157820204

[GOL01] Goldstein, Larry J., Schneider, David I., Siegel, Martha J.; Finite Mathematics and Its Applications, Fourth Edition, Prentice Hall, 1980, ISBN 0-13-318221-5

[HIG001] Higham, Charles; Trading With the Enemy: An Expose of the Nazi-American Money-Plot 1933-1949, Hale, London 1983

[HOD01] Hodges, M. Susan; Computers: Systems, Terms and Acronyms, 16th Edition; SemCo, 2006, ISBN: 9780966842289

[HUF01] A Method for the Construction of Minimum redundancy Codes, David A. Huffman, Proceedings of the I.R.E., Volume 40, Issue 9, Sept. 1952, pgs 1098-1102, ISBN 0096-8390

[JEN01] Jennings, Tom; An Annotated History of Some Character Codes or ASCII Infiltration; http://www.wps.com/projects/codes/

[JOH001] Johnson, Dr. Neil F., www.jjtc.com/ Steganography

[KAH01] Kahn, David; The Codebreakers: The Story of Secret Writing; Scribner, New York, 1996; ISBN 0684831309

[KAH02] ibid. pg. 523

[KAH03] ibid. Chapter 16

[KAH04] ibid. pg. 518

[KAH05] ibid. Chapter 20

[KAH06] ibid. pgs. xv-xviii

[KAH07] ibid. pg. 508

[KAH08] ibid. Chapter 17

[KAH09] ibid. pg. 100

[KAH10] ibid. pgs. 739-740

[KAH11] ibid. pg. 737

[KAH12] ibid. pgs. 100-105

[KAH13] ibid. Chapter 4

[KAH14] ibid. pg. 488

[KAH15] ibid. pg. 748

[KNO01] Knott, Ron; Using the Fibonacci Numbers to Represent Whole Numbers, http://www.mcs.surrey.ac.uk/Personal/R.Knott/Fibonacci/fibrep.html

[LIV01] Livio, Mario; The Golden Ratio: The Story of Phi, The World's Most Astonishing Number; Broadway Books, New York, 2002, ISBN: 0-7679-0815-5; pg. 11

[LIV02] Ibid. Chapter 5

[LIV03] Ibid. pgs. 232-236

[LIV04] Ibid. Appendix 9

[MAL01] Digital Principles and Applications, Fourth Edition, Malvino, Albert Paul and Leach, Donald P.; Macmillan/McGraw Hill, 1986, ISBN 0-07-039883-6

[MIA01] Compressed Image File Formats: JPEG, PNG, GIF, XBM, BMP, Your guide to graphics; by John Miano, ACM Press, SIGGRAPH Series, 1999

[PEN01] Pennebaker, William B. and Mitchell, Joan L.; *JPEG Still Image Data Compression Standard*, Van Nostrand Reinhold, 1993

[PHI01] Philologos; Boydem II:Yiddish and Cockney? The Jewish Daily Forward; Friday, December 22, 2006;

[VAU01] Vaudenay, Serge; A Classical Introduction to Cryptography: Applications for Communications Security; Springer Science+Business Media, Inc., New York, 2006, ISBN-10: 0-387-25464-1

[VAU02] Vaudenay, Serge; A Classical Introduction to Cryptography: Applications for Communications Security; Springer Science+Business Media, Inc., New York, 2006, ISBN-10: 0-387-25464-1; Section 1.1.2, definition of decryption

[VAU03] Vaudenay, Serge; A Classical Introduction to Cryptography: Applications for Communications Security; Springer Science+Business Media, Inc., New York, 2006, ISBN-10: 0-387-25464-1; Section 1.1.3

[VER01] Verne, Jules; Twenty Thousand Leagues Under the Sea, Sterling Publishers, New York, 2006, ISBN: 140272599X

[WEB01] Webster's New International Dictionary, Second Edition, G & C. Merriam Co., Springfield, Massechusetts; 1945; entry for cipher

[WEB02] Webster's New International Dictionary, Second Edition, G & C. Merriam Co., Springfield, Massechusetts; 1945; entry for code

[WEB03] Webster's New International Dictionary, Second Edition, G & C. Merriam Co., Springfield, Massechusetts; 1945; entry for argot

[WHO01] Language, Thought and Reality: Selected Writings of Benjamin Lee Whorf, MIT Press, Cambridge Massachusetts; 1993; ISBN 0-262-2303-8

[WIK01] Golden ratio base; http://en.wikipedia.org/wiki/Phinary#Writing_golden_ratio_base_numbers_in_standard_form

[WIK02] Non-standard Positional Numeral Systems, http://en.wikipedia.org/wiki/Non-standard_positional_numeral_systems

[WIK03] Endianness http://en.wikipedia.org/wiki/Big_endian

[WIK04] Encryption; http://en.wiktionary.org/wiki/encryption

[WIK05] CAPTCHA; http://en.wikipedia.org/wiki/Captcha

[WIK06] Golden ratio; http://en.wikipedia.org/wiki/Golden_ratio

[WIK07] Fibonacci Number; http://en.wikipedia.org/wiki/Fibonacci_number

[WIK08] Fibonacci coding; http://en.wikipedia.org/wiki/Fibonacci_representation

[WIK09] Zeckendorf's theorem; http://en.wikipedia.org/wiki/Zeckendorf%27s_theorem

[WIK10] International Phonetic Alphabet, http://en.wikipedia.org/wiki/International_Phonetic_Alphabet

[WIK11] IPA Chart for English, http://en.wikipedia.org/wiki/International_Phonetic_Alphabet_for_English

[ZIR01] Zirkind, Givon; AFIS Data Compression: An Example of How Domain Specific Compression Algorithms Can Produce Very High Compression Ratios; ACM SIGSOFT, Software Engineering Notes, Volume 32, Issue 6, November 2007, Article No. 8, ISSN: 0163-5948

# GLOSSARY

Cipher or Encipher – A cipher is a set of symbols or letters used to replace intended letters to create a secret message. Enciphering is the process of substituting the letters of one message with a cipher. [WEB01] Commonly, especially with computers, numbers are used for letters and; the numbers are interchanged. I.e. A=1, B=2, C=3, etc. Ciphering might be as simple as adding one to each number. Or, ciphering might involve a more complex mathematical operation.

Code or Encoding – is to translate into symbols. [WEB02] Using some kind of one to one correspondence, a translation of symbols is made. Letters and words are both symbols and; can both be encoded.

Enciphering – is a special kind of encoding, when substituting one set of symbols (letters) for another set of (letters).

Encryption – is new term for scrambling information or data. Encryption is not limited to letters or words. Also, encryption is not limited to substitution or a one to one correspondence, such as a codebook or using a key per se. Commonly, encryption involves complex algorithms, usually employing complex mathematical formulae. [WIK004] [KAH05] [KAH06]

Deciphering – is to reverse the process of enciphering.

Decoding – is to reverse the process of encoding.

Decrypting – is to reverse the process of encrypting. However, technically or commonly, decrypting connotes reversing the encryption when NOT in possession of the decryption key. [VAU01]

Apart from the above methods of making messages secret and keeping communications secure; there is yet another way of keeping messages secret: That is to speak another, not understood, "secret," language. This practice has been used by thieves, the military, spies and private investigators. Immigrant parents often use their native tongue as "the secret" language. Secret languages are also a common tactic used in price negotiations in markets. Argot is an example of this. [PHI01] [WEB03]

Boustrophedon – is when a language is written from either right to left or; left to right. This was true of some ancient languages, including Greek, up until a certain time. [KAH001] [FRI001]

Little Endian / Big Endian – This refers to which digit, the right or the left, is the biggest, or most significant digit. [BET01] [BLA01] [WIK03] Little Endian systems have the smallest digit on the right. While Big Endian systems, have the biggest digit on the right. Each successive digit, is an additional multiple of the base. E.g. In base 10: $10^2$ x $10^1$ x $10^0$. E.g. In binary numbers: "0001" could be a decimal 1. Or, "1000" could be a decimal 1. Or, "1000" could be a decimal 8 [2x2x2]. It depends where you put the big end and where you put the little end. This has a bearing upon how binary numbers are written, represented and actually placed onto hardware. While little endian is common in writing, i.e. "0001" is a decimal 1; machines may actually operate in big endian format.

In terms of encryption, one can consider reversing big endian with little endian, like boustrophedon or mirror writing. Meaning, to write the letters of a message backwards.

CAPTCHA – Completely Automated Public Turing test to tell Computers and Humans Apart. A test, given as a challenge, by a computer, that a computer can not answer. If the question is answered correctly, it is assumed the respondent is human. This test usually involves reading a distorted image. [CAP01] [WIK05]

Binary Code – A code with only two symbols. E.g. on/off, 1/0, etc. These two symbols can be combined in any fashion to encode any number of things. Morse code is one example of a binary code. Base 2 is another common well known example of a binary code.

The Golden Ratio – is the ratio that satisfies the ratio a:b::(a+b):a. Or, a/b = b/(a+b). This equation becomes $a^2+ab-b^2=0$. Using the quadratic equation, the equation can be solved. The solution to the equation is a constant, equal to $(1 + \sqrt{5})/2 \approx 1.6$  The golden ratio is symbolized with the Greek letter Phi: φ  The golden ratio has many unique properties that have made it the object of study. [EUC01] [EUC02] [LIV001] [WIK06]

The Golden Mean – same thing as the golden ratio. Although, the golden mean is more of a geometric property, rather than an algebraic description of the ratio.

The Fibonacci Sequence – is a sequence named after the mathematician, Leonardo Fibonacci, who did not discover this sequence, but wrote about it, in his book, Book of Calculation, the Liber Abaci. [LIV001] [KNO01] The sequence starts at zero. Followed by a one. Thereafter, each consecutive number is the sum of the previous two consecutive numbers: {0, 1, 1, 2, 3, 5, 8, 13… }  [LIV02] [KNO01] [WIK07]

A Fibonacci number – is a number in the Fibonacci sequence. [LIV02] [KNO01] [WIK07]

The Fibonacci sequence has many unique properties.  One property is, that any whole number, positive integer, less than any given Fibonacci number, can be expressed as the sum of some subset of the preceding Fibonacci numbers. E.g. Using the above subset of Fibonacci numbers, one can count up to 13 thus: 1, 2, 3, 1+3, 5, 1+5 or 1+2+3, 2+5, 8, 1+8, 2+8 or 1+1+8, 3+8 or 1+2+8, 1+3+8 or 1+1+2+8, 13. [LIV02] [KNO01] [WIK07]

Fibonacci Representation – is a numeric representation that uses the property of the Fibonacci series that permits the expression of any positive integer as a sum of Fibonacci numbers. This numeric representation is a natural extension of the Fibonacci series; making Fibonacci representation a natural binary system. [LIV01] [KNO01] [WIK01]  I.e. Using a left to right sequence, big endian, system, and a zero for exclusion, and a one for inclusion: "$01111_{FIB}$" = (in decimal) $0+1+1+2+3$ = $7_{10}$   Even though the Fibonacci is a natural binary system, the mathematics are more complicated than base 2.

Standard Form Fibonacci Representation – In addition, the Fibonacci sequence has a property, proven by Zeckendorf's theorem, that any number can be represented by a set of previous Fibonacci numbers, without any two Fibonacci numbers in a row. E.g. Four in base 10, could be "$0111_{FIB}$" or "$01001_{FIB}$" or "$00101_{FIB}$". [KNO01] [WIK09]

Standardized Form – is when, in a binary system, for a given type of number, e.g. positive integer, every number can be expressed without consecutive ones. E.g. "11" is not present in the numbering system. [WIK09]   When applying Zeckendorf's theorem [see previous paragraph] to the Fibonacci based numbering system, then, the Fibonacci numbers are being expressed in standardized form.

Golden Ratio Base – Using the golden ratio as a base, any real number can be expressed as a binary number. This is also referred to as Phinary, after the name Phi, φ, for the golden ratio. Numbers in Phinary are written thus: $0101_φ$   In addition, like Fibonacci numbers, any golden ratio base number can be written in a standardized form. [WIK01] [WIK06] [WIK09]

Golden Sequence – a binary sequence, a long range numeric sequence, that is not periodic; based upon the Fibonacci sequence. The sequence is generated by starting with a "1". Then, replacing each "1" with a "10" and, each "0" by a "1":

1
10
101
10110
10110101
1011010110110

Each sequence is a combination of the last two previous sequences – in Fibonacci fashion. The sequence is "self-similar" and expandable infinitely; with uniqueness.

Cleartext or Plaintext – Regular text that has not been enciphered, encoded or encrypted. [KAH06] [VAU01]

While I have mentioned only four natural binary systems: Fibonacci and Phinary, both standard and non-stardard; there are many more binary systems that could be constructed, as will be discussed below.

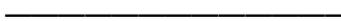